\documentclass[a4paper,11pt]{article}
\pdfoutput=1 % if your are submitting a pdflatex (i.e. if you have
\usepackage{jinstpub} % for details on the use of the package, please
\usepackage{lineno}
\title{SPY: A Conceptual Design Study of a Magnet System for a High-pressure Gaseous TPC Neutrino Detector}

\author[1]{Andrea Bersani}
\author[2]{Alan D. Bross}
\author[2]{Michael Crisler}
\author[1]{Stefania Farinon}
\author[3]{Christopher Hayes}
\author[2]{Donald Mitchell}
\author[1]{Riccardo Musenich}
\author[2]{Colin Narug}
\author[2]{Jay Theilacker} 
\author[2]{Terry Tope }
\author[2]{Erik Voirin}
\author[4]{Vivek Jain}
\affiliation[1]{Istituto Nazionale di Fisica Nucleare, Genoa, Italy}
\affiliation[2]{Fermi National Accelerator Laboratory, Batavia, IL, USA}
\affiliation[3]{Indiana University, Bloomington, IN, USA}
\affiliation[4]{University of Albany, SUNY, Albany, NY, USA}
\emailAdd{bross@fnal.gov}

\abstract{
We present a novel conceptual design for a magnet system that provides the magnetic field necessary for the analysis of tracks in a high-pressure gaseous argon TPC while simultaneously serving as a pressure vessel to contain the TPC gas volume.  The magnet was developed within a Near Detector proposal for the Deep Underground Neutrino Experiment (DUNE). The high-pressure gaseous argon TPC is a component proposed to be one of the elements of an ensemble of near detectors that are needed for DUNE.} 

\arxivnumber{2311.16063} % only if you have one

\collaboration[c]{on behalf of the DUNE collaboration}

\usepackage[utf8]{inputenc}
\usepackage{graphicx}
\usepackage{amsmath}
\usepackage{xcolor}
\usepackage{epsfig}
\usepackage{url}
\usepackage{authblk}
\usepackage{lineno}

\usepackage{hyperref}
\usepackage{color}
\usepackage{fancyhdr}
\fancypagestyle{magheader}{
\fancyhf{}
\rhead{\em ND-GAr Magnet System}
\fancyfoot[C]{\thepage}
\renewcommand{\headrulewidth}{2pt}
\renewcommand{\headrule}{\hbox to\headwidth{\color{blue}\leaders\hrule height \headrulewidth\hfill}}
}
\usepackage{natbib}
\usepackage{graphicx}

\newcommand{\unit}[1]{\,\mathrm{#1}}

\usepackage{subcaption}

\begin{document}
%\flushbottom
%
%
%
%
\maketitle
\section{Introduction}
\label{Intro}
A key aim of the DUNE experiment is to measure neutrino interaction rates from which the oscillation probabilities for muon (anti)neutrinos to either remain the same flavor or oscillate to electron (anti)neutrinos can be extracted.  The DUNE Far Detector, located at the Sanford Underground Research Facility, $1300\unit{km}$ away from the neutrino source at Fermilab, will measure the neutrino interaction rate after oscillations.  The Near Detector complex, located on the Fermilab site $\simeq570\unit{m}$ from the neutrino target, will measure the un-oscillated neutrino flux, providing the experiment’s control sample.
A robust understanding of the neutrino flux at the source will require measurements both on and off the beam axis at the near site, in addition to continuous monitoring of the on-axis flux which will be done by a beam monitor called SAND (System for on-Axis Neutrino Detection). 

\begin{figure}[h!]
\centering
\includegraphics[width=0.98\textwidth]{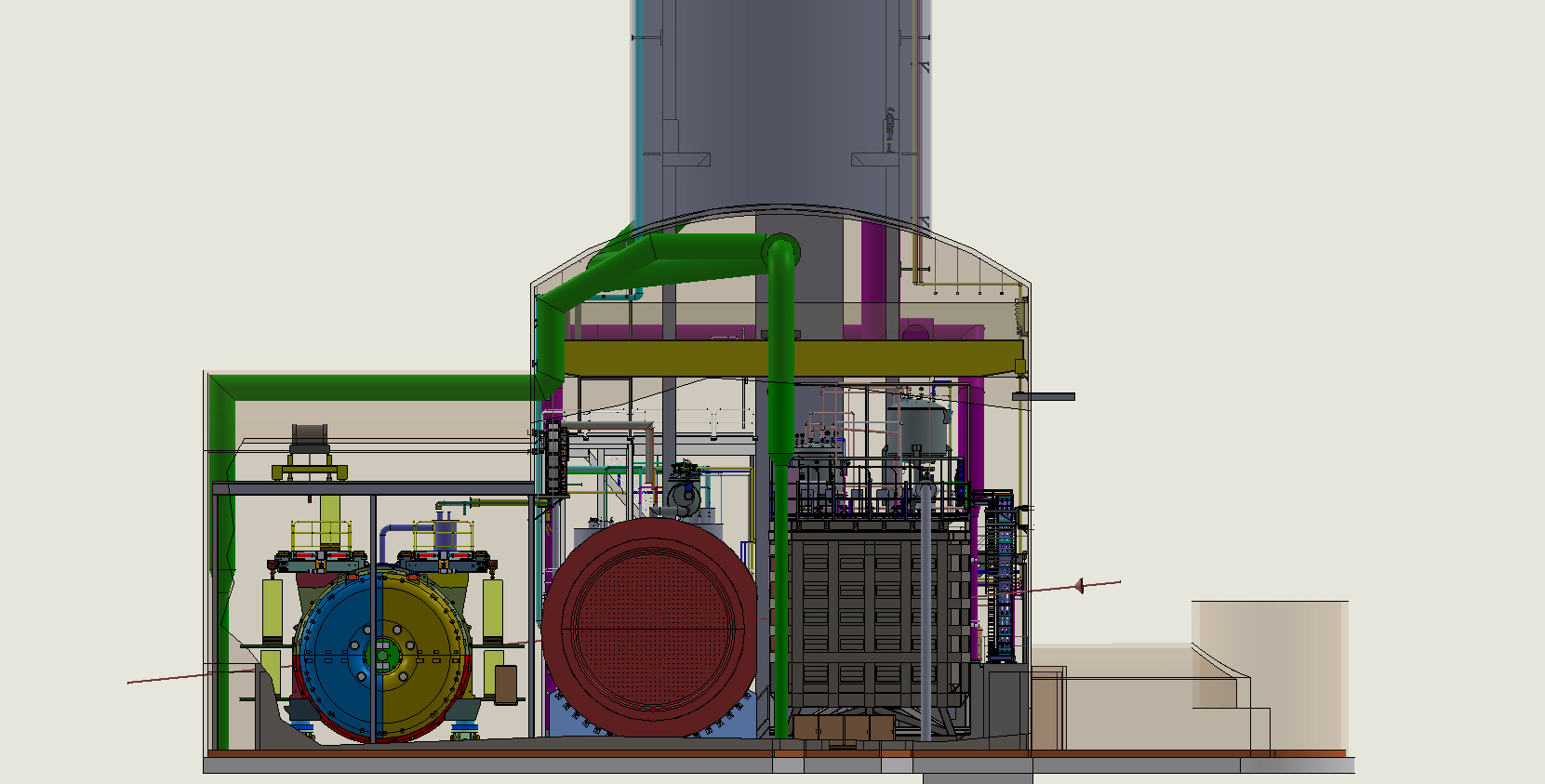}
\caption{The Near Detector Suite of Detectors: the neutrino beam travels from the right to the left, crossing in sequence ND-LAr, ND-GAr and SAND}
\label{fig:ND}
\end{figure}

Detailed studies of the neutrino flux both on and off axis will be be done using a modular liquid argon  detector with pixel readout called ND-LAr, supplemented  by an iron range stack to measure the momentum of muons exiting ND-LAr.  This will be the initial configuration.  In order to meet all of DUNE's physics goals, a detector that can measure neutrino interactions on argon with a precision even better than in ND-LAr is needed, however.  This detector must also measure the momentum of muons that exit ND-LAr as mentioned above.  A proposal for this enhanced Near Detector, called ND-GAr, includes a high-pressure (10 bar) gaseous argon time projection chamber (HPgTPC) system~\cite{Mart_n_Albo_2017} surrounded by an electromagnetic calorimeter where both are in a magnetic field.    ND-GAr must also be designed to be movable and to operate in multiple positions.   The movable components of the near detector (ND-LAr + TMS and then ND-LAr + ND-GAr) will move approximately 30 m off axis which corresponds to approximately 3$^\circ$.  The detectors will be mounted on 200 t capacity motorized Hilman skates~\cite{site:Hilman}.  See Figure~\ref{fig:Stayed-head}.  Cryogenics, power, communications are provided via a flexible energy chain design~\cite{site:EChain}.  The complete DUNE Near Detector system including ND-GAr is shown in Figure~\ref{fig:ND}.  In order to maximize muon acceptance, the distance between ND-LAr and ND-GAr (active to active) is kept to a minimum and is approximately 3.5m.

Magnet options that were considered for ND-GAr included a room temperature dipole similar to the UA1 magnet and 5 superconducting coils in a Helmholtz configuration~\cite{Bross:2019ynq,instruments5040031}. We eventually became aware of a magnet 
produced by ASG Superconductors in Italy for the Multi Purpose Detector (MPD) at the NICA Collider at JINR~\cite{MPD-JINR}. The MPD-JINR magnet has a smaller bore than is required for ND-GAr, but otherwise meets all the requirements for the ND-GAr magnet. In addition ASG felt that they could deliver a solenoid magnet suitable for ND-GAr at a cost significantly below our estimated cost for the Helmholtz coil system.  The magnet group within ND-GAr then focused on a solenoid solution which has now become the baseline design.  The ND-GAr solenoid design closely follows the concepts developed for the MPD-JINR magnet.

\begin{figure}[h!]
\centering
\includegraphics[width=0.65\textwidth]{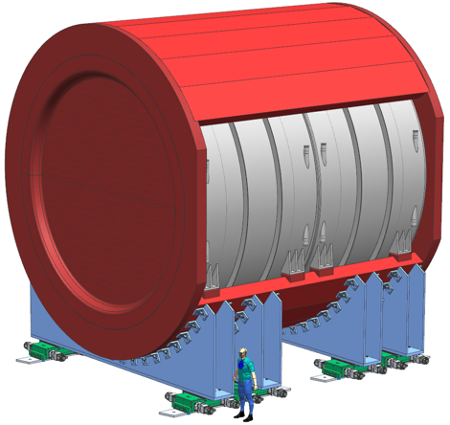}
\caption{The SPY magnet system.}.  
\label{fig:SPY}
\end{figure}

\begin{figure}[b!]
\centering
\includegraphics[width=0.98\textwidth]{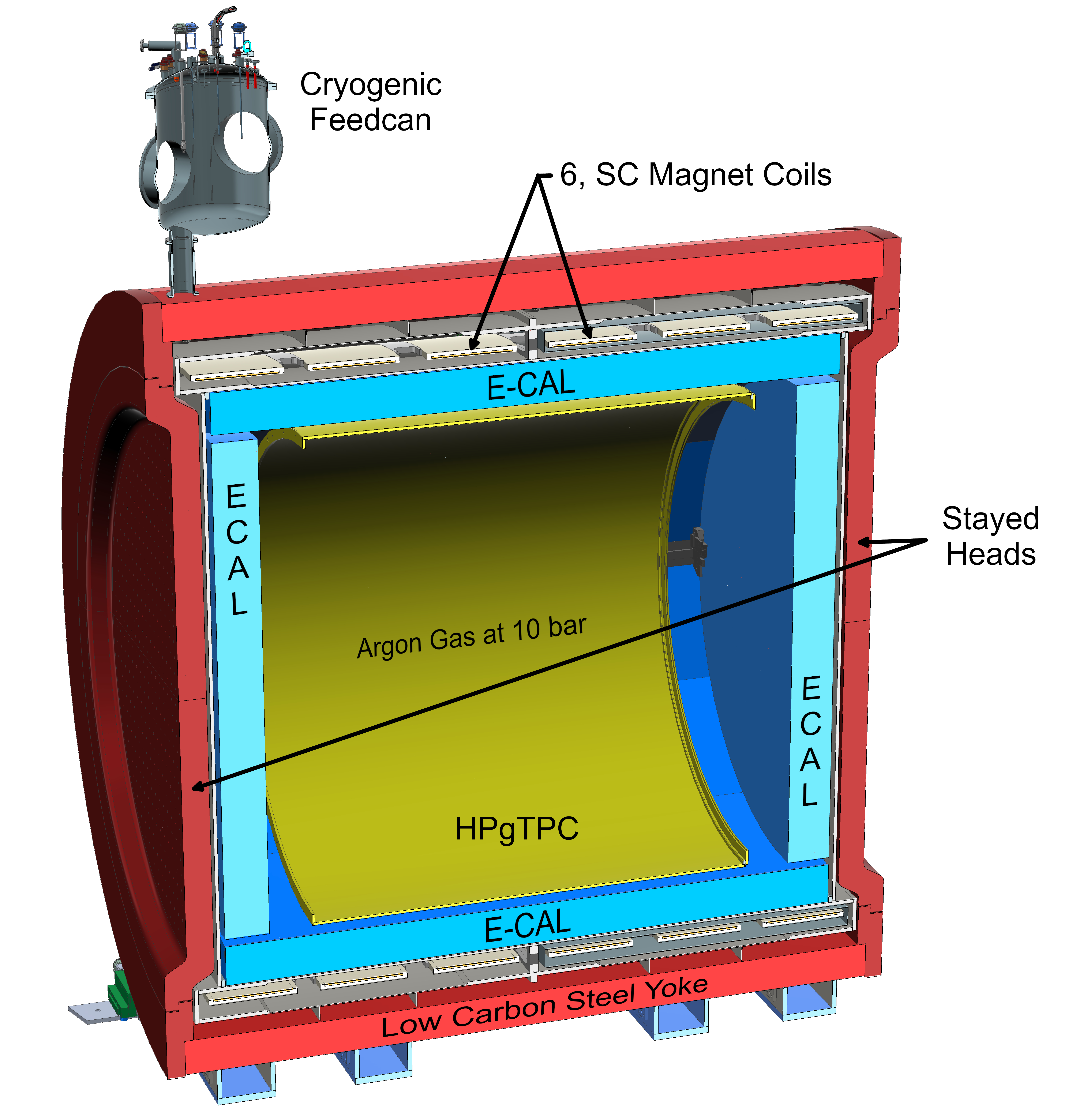}
\caption{Cut-away view showing the various components of ND-GAr.}  
\label{fig:SPY_c}
\end{figure}

The focus of this paper is the conceptual design of an integrated magnet and pressure vessel system for ND-GAr.  The magnet system consists of a superconducting solenoid surrounded by an iron return yoke. To control the physical size and cost of the magnet system, we have developed an integrated design for the superconducting solenoid cryostat so that it will also serve as the cylindrical component of the pressure vessel for the HPgTPC, while at the same time providing support for the HPgTPC and calorimeter elements located in its bore.  The mechanical design and stress analysis of the solenoid cryostat will be presented in subsequent sections.
Additionally, the design of the iron magnet yoke uses the mechanical strength of the yoke's pole faces to eliminate the large domed heads that would normally be required for a large-diameter pressure vessel.  The stayed-head design shortens the overall dimension of the system transverse to the beam by approximately 4m.  The incremental cost of strengthening the solenoid cryostat is small compared to the cost of a separate pressure vessel which is estimated to be greater that half the cost of the superconducting solenoid.  The stayed-head design that closes the pressure vessel will be described in detail in Section 6.
An important design requirement for ND-GAr is the ability to accurately measure the momentum of muons that originated in ND-LAr.  This requirement limits the amount of material allowed on the upstream side of ND-GAr and forces us to adopt an unsymmetrical iron yoke.
To address this issue, we have developed an iron yoke that eliminates a portion of the iron along the entering particle paths. The system is called SPY -- Solenoid with Partial return Yoke.   A schematic of ND-GAr is shown in Figure~\ref{fig:SPY} where the missing section of the yoke is shown.  A cut-away view is shown in Figure~\ref{fig:SPY_c} which shows the coils, ECAL components, and the
HPgTPC.  A possible location for the cryogenic feed can is also shown. Development of the design concepts for the HPgTPC and the ECAL continue, but the overall dimensions and requirements have been defined in an earlier phase of the proposal's development~\cite{instruments5040031}, allowing for a reliable design of the magnet system.

%The experiment also considered a temporary initial configuration of ND-GAr that does not include the HPgTPC or the ECAL (to be installed at a later time) that would only have a scintillator tracker inside the magnet bore.  This configuration is called ND-GAr-Lite (see Figure~\ref{fig:nd-gar-lite}) {\textcolor{red}{Should we ref ND-GAr-lite CDR?}}.
%
%
%
%
%\begin{figure}[ht!]
%\centering
%\caption{ND-GAr-Lite with the end iron removed showing the scintillator %tracker planes}
%\label{fig:nd-gar-lite}
%\end{figure}
In this document we will layout the requirements for the SPY magnet system and present details of the conceptual design.  The superconducting solenoid design is conservative, following the design of existing magnets and using known best practices in the field of superconducting magnet technology.  We have analyzed the impact of the partial yoke design on magnetic field uniformity and on fringe fields and show that both can meet the requirements of the experiment.
\section{Magnetic system specifications}
As discussed in Section~\ref{Intro}, ND-GAr is one component of a potential three-component suite of detectors at the DUNE near site.  As such, the magnetic system has to meet ND-GAr specifications and has to operate within a number of requirements imposed by the 3-detector configuration.  They are:
%https://www.overleaf.com/project/61ead3ebd5f5fb4d1d40a2c0
\begin{itemize}
    \item {Magnetic: The momentum analyzing power of ND-GAr must provide at least 3\% momentum resolution for the ND-LAr muons.  For particles produced as a result of neutrino interactions in the argon gas, the analyzing power must produce a resolution in neutrino energy reconstruction at least as good as that of the DUNE far detector }
    \item {Geometrical: ND-GAr must be able to measure the momentum of muons exiting ND-LAr.  Monte Carlo simulations have shown that the HPgTPC baseline (active volume 5 m in diameter and 5 m long) will have acceptance to meet DUNE's goals for the near detector~\cite{DUNE:2022yni}.  The HPgTPC drives the size of the magnet.  With the addition of the ECAL, a bore of approximately 7 m is required.  In addition, in order to have good acceptance for low-energy muons when the detector is positioned far off axis and to not degrade muon momentum resolution, the amount of dead material in the muon's path must be minimized.  The partial return yoke accomplishes this.  See Section~\ref{sec:MB}.}
    \item {Mechanical: ND-GAr's magnet system must have a minimum quantity of material in the downstream face of the yoke to assist in the discrimination of muons from pions.  A symmetric yoke (window both upstream and downstream) is therefore not an option.  In addition a design with a thin (in g/cm$^2$) solenoid is required.  A symmetric magnetic configuration is possible if the downstream material is non-magnetic (stainless steel for example), but this increases the cost and produces stray field issues. }
\end{itemize}
In this section we will describe in detail the SPY magnet system design specifications.

\subsection{Magnetic: Field and field quality}

The main requirement for the SPY magnet system is on the magnetic field that will be needed by the tracker that will be used in this detector.  The HPgTPC design is based on the ALICE TPC~\cite{Alme:2010ke} at the LHC. The TPC will be cylindrical, $\simeq 5.2\unit{m}$ long and $\simeq 5.2\unit{m}$ in diameter. The TPC axis will be horizontal and perpendicular to the neutrino beam direction.  The HPgTPC will provide excellent tracking resolution and we have determined that a relatively low magnetic field of $0.5\unit{T}$ will be sufficient to attain the desired momentum resolution.

Thanks to the recent and expected future improvements in software reconstruction and computing power, the requirement on field uniformity is significantly looser than in previous TPC-based detectors. From this perspective, the requirement is $\pm10\%$ with the stipulation that an accurate field map of the ``as-built" system is performed.  The field quality achieved in the simulation of our current magnet system design already significantly exceeds this specification (see Table~\ref{tab:fq} below).

\subsection{Geometrical constraints}
The outer size of the magnet system is constrained by the available space in the experimental hall. The maximum height is defined by the $12\unit{m}$ clearance under the overhead crane. This is not a real constraint for the magnetic design, but it may impact the design of the cryogenic feed can (see Figure~\ref{fig:SPY_c}). The width of the iron return yoke, in the beam direction, is the most constrained dimension as shown in Figure~\ref{fig:ND}. The available space is $8.82\unit{m}$, in which a stay-clear between ND-GAr and ND-LAr on one side and between ND-GAr and the wall on the other side is required.
To achieve the best utilization of the available space, a novel integration approach has been developed, which uses the solenoid cryostat as the HPgTPC pressure vessel body and uses the mechanical strength of the magnet yoke to close the pressure vessel ends with very thin covers using a stayed head design.

\subsection{Mechanical: Material budget}
\label{sec:MB}
An important systematic uncertainty on the measurement of muons that exit the LAr detector arises from muon energy loss in non-active material between ND-LAr and ND-GAr.  In order to determine the muon momentum with the required precision, we have imposed a requirement, based on simulations, that the total amount of dead material in the muon path as it travels from the active region of ND-LAr to the active region of ND-GAr be less than 100 g/cm$^2$.  The dead material in ND-GAr is limited to 50\% of this amount, or 50 g/cm$^2$.  The opening in the return yoke solves this problem for the iron, so this requirement defines the total mass allowed for the solenoid and its cryostat/pressure vessel.  The material budget for the current solenoid design is shown in Table~\ref{tab:MB}.  The reported values already consider some contingency, namely on the coil former thickness, and therefore we can conclude that the design fulfills the 50 g/cm$^2$ limit.

\begin{table}[h]
\small
\begin{center}
\caption{Material budget for solenoid.}
\label{tab:MB}
\begin{tabular}{|l|c|c|c|r|}
\hline
Component &  Material & Thickness (mm) & $\rho$ (g/cm$^3$) & g/cm$^2$ \\ \hline\hline
Outer vacuum vessel wall & Stainless steel & 25 & 7.87 & 19.68 \\ \hline
LN$_2$ shield & Al & 4 & 2.70 & 1.08 \\ \hline
Coil former & Al & 40 & 2.70 & 10.8 \\ \hline
Coil & mainly Al & 20 & 2.70 & 5.4 \\ \hline
LN$_2$ shield & Al & 4 & 2.70 & 1.08 \\ \hline
Inner vacuum vessel wall & Stainless steel & 16 & 7.87 & 12.59 \\ \hline
\textbf{TOTAL} & & & & \textbf{50.63} \\ \hline
\end{tabular}
\end{center}
\end{table}
%
%\newpage https://www.overleaf.com/project/607982233fdc7889d7b2b34d
%\subsection{Loading (DM, CN)}
%\paragraph{HPgTPC}
%The HPgTPC weight of $\simeq$ 14t is supported by 2 rails that are attached to the inner surface of the ECAL.
%\paragraph{ECAL}
%The ECAL weight of $\simeq$ 150t is heavy. It is installed in wedge-shaped segments to the inner diameter surface of the pressure vessel.
%
\section{Magnetic design}

\subsection{Design principles}

%Several different layouts have been studied in order to meet the complexity of coping with the strict geometrical requirements for ND-GAr. Among the proposed solutions, a concept using a ``Helmholtz-like'' five coil configuration~\cite{Bross:2019ynq} was developed. However, a new concept based on a continuous thin solenoid is now regarded as the baseline design, the main advantages over the previous design being the management of the stray field and the field quality.
The solenoid design is based on the decades-long evolution of internally wound, aluminium-stabilised cable for superconducting magnets, starting with CELLO~\cite{Desportes:1979mf}, and including CDF~\cite{Fast:1982qt}, Delphi~\cite{Clee:1989iq}, BaBar~\cite{Fabbricatore:1996mc}, and many others. The $0.5\unit{T}$ central field permits a single-layer coil to provide the needed current density even with our very large magnetic volume. The design parameters are conservative when compared to previously built magnets.  The main parameters of the proposed magnetic design are summarized in Table~\ref{tab:fq}.

The magnetic calculations have been performed with ANSYS Maxwell finite element software which is a state-of-art optimized tool for the simulation of low-frequency electromagnetic fields in industrial components. It includes 3D/2D magnetic transient, AC electromagnetic, magnetostatic, electrostatic, DC conduction and electric transient solvers to accurately solve for field parameters including force, torque, capacitance, inductance, resistance and impedance~\cite{site:maxwell}.
\begin{table}[h]
\small
\begin{center}
\caption{SPY magnetic parameters.}
\label{tab:fq}
\begin{tabular}{|l|c|c|}
\hline
\textbf{Parameter} & \textbf{Value} & \textbf{Unit }\\ \hline \hline
Central field & 0.5 & T \\ \hline
Field uniformity & $\pm1$ & $\%$ \\ \hline
Operating current & 4585 & A \\ \hline
Inductance & 2.75 & H \\ \hline
Force on coil (in neutrino beam direction) & 150 & kN \\ \hline
SAND magnet stray field in ND-GAr tracker & $<10^{-3}$ & T \\ \hline
SPY stray field in SAND tracker & $<5\cdot10^{-4}$ & T \\ \hline
SPY stray field in ND-LAr fiducial volume & $<10^{-2}$ & T \\ \hline
\end{tabular}
\end{center}
\end{table}

\subsection{Coil and coil former design}
\label{sec:Coil}

The coil design is based on a rectangular cable with dimensions 
$\simeq 20 \times 7.5$ mm$^2$ and will be wound on its long axis, the so-called
``hard-way bend" wind.  With this cross section, the overall current density is $\sim
30.5\unit{A/mm^2}$. An analysis of the benefits of a reduction of the inductance (fewer turns/higher current), to
allow for faster charge and discharge of the magnet, versus more turns with lower current, to keep 
the voltage as low as possible during quenches drove this choice.
The maximum field on the cable, according to our
calculations, is below $1\unit{T}$. The cable supplier will be requested to
supply a cable with a sufficient amount of superconductor such that the cable
can carry twice the design current at twice the maximum field at the operating
temperature, i.e. $10,000\unit{A}$ at $2\unit{T}$ and at $4.5\unit{K}$. The
superconductor will be co-extruded in high purity aluminium to provide quench
protection in the worst case. A possible solution for the cable, based on Niobium
Titanium, could be a Rutherford cable made of 10 strands, $0.8\unit{mm}$
diameter, 1:1 Cu/SC ratio, co-extruded in high purity aluminum.
\begin{figure}[b!]
\centering
\includegraphics[width=0.8\textwidth]{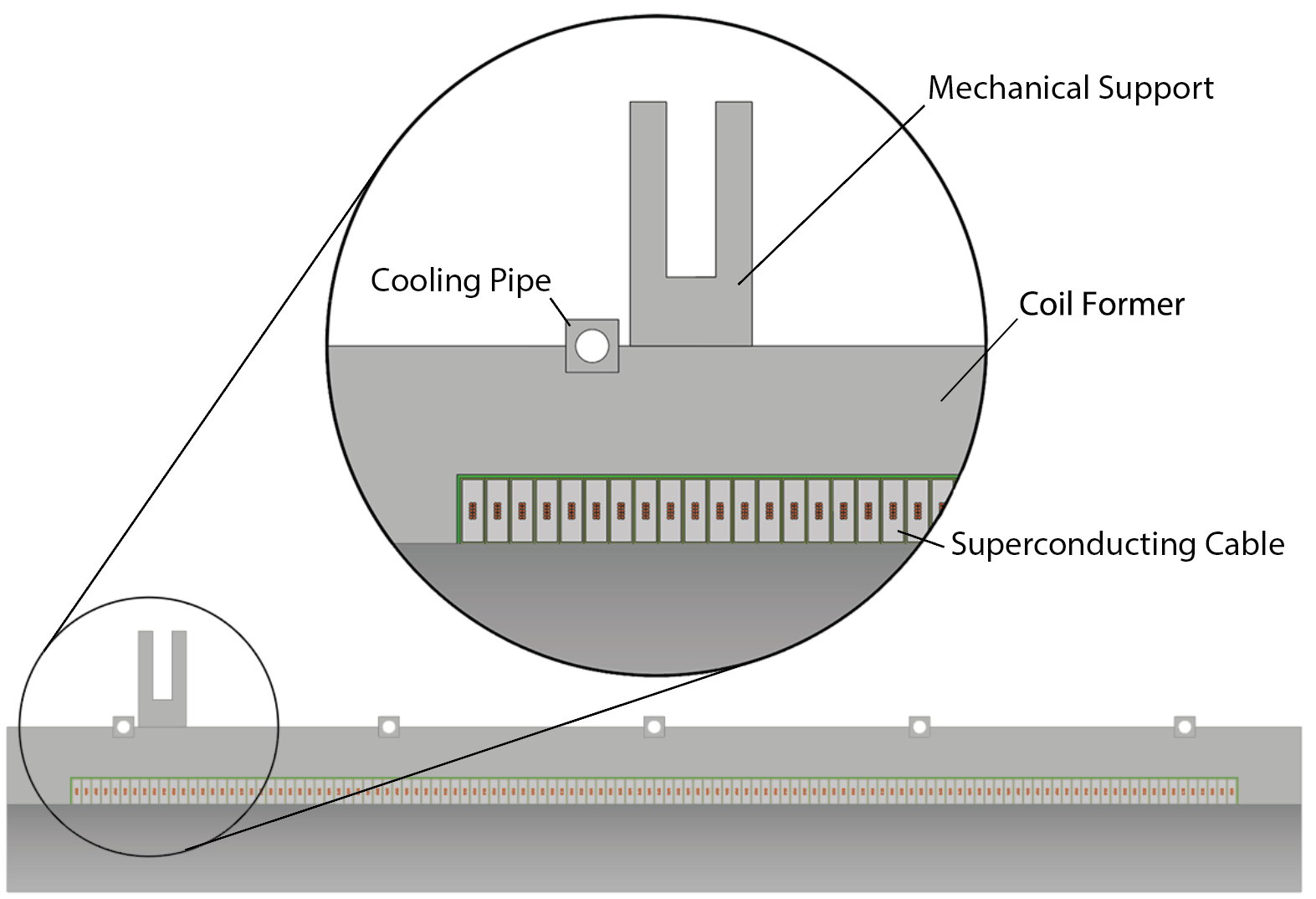}
\caption{Conceptual cut-away view showing the coil in the coil former with the main components of the cold mass. The complete solenoid will be built out of six coil assemblies (coil + former), connected with spacers.}  
\label{fig:bobbin}
\end{figure}
The coil will be built in segments (see Figure~\ref{fig:SPY_c}), to be
joined before insertion in the cryostat. Six identical subcoils are
foreseen, each with a $7000\unit{mm}$ internal diameter, $900\unit{mm}$ length
and $20\unit{mm}$ thickness. Each subcoil will be internally wound in a coil
former made of aluminium alloy. The subcoils will then be mechanically joined
with spacers and the electrical connections between the superconducting cables
will be made. Each subcoil will provide $550\unit{kA\cdot turn}$, for a total of
$3.3\unit{MA\cdot turn}$. As a design guideline, we decided to keep the current
below $5000\unit{A}$ to avoid high voltages during quenches. This can be
achieved with 120~turns for each subcoil operating at $\sim 4585\unit{A}$. The
calculated stored energy with this configuration is $\sim 32.5\unit{MJ}$. The
inductance of the magnet is $\sim 2.75\unit{H}$.  The coil dimensions are summarized
in Table~\ref{tab:CD}. 
\begin{table}[h]
\small
\begin{center}
\caption{Coil Dimensions.}
\label{tab:CD}
\begin{tabular}{|l|c|c|}
\hline
Measurement & Former (mm) & Coil (mm) \\ \hline
ID & 7000 & 7000 \\ \hline
OD & 7120 & 7040 \\ \hline
Length & 1000 & 900 \\ \hline
\end{tabular}
\end{center}
\end{table}

\subsection{Thermal design philosophy and cryogenic delivery system}
  In order to reach an operating temperature of between 4.5K and 4.7K, the six superconducting coils will be conduction cooled by a thermosiphon-driven flow of liquid helium in pipes welded onto the outer surface of the coil former as shown in
Figure~\ref{fig:bobbin}.  An aluminum thermal shield is also required for stable operation and to minimize heat load to the liquid helium.  It will be either cooled by cold helium gas or by liquid nitrogen, depending upon the final refrigerator design. This implies that the shield would operate either near 50K or near 80K. The feed can design is only conceptual at this point, but we have chosen high-temperature superconductor for the current leads to minimize the liquefaction load on the cryogenic refrigerator and the thermal load in general. 
  
  The cryogenic fluids will be provided through a cryogenic distribution system in the experimental hall and will deliver liquid helium and liquid nitrogen in vacuum insulated flex-hoses supported by an articulating pipe carrier.  The system provides cryogens to the feed-can that will be mounted on a work platform that is secured to the top of SPY. 
  
  These cryogenic services will be installed in parallel during the construction of ND-GAr.  The feed-can installation followed by cryogenic connections and coil lead splices will be the last activities necessary to complete the cryo system.   We note that the superconducting magnet assembly will have already been tested at the vendor fabrication site for vacuum leaks, cryosgenic issues at 4.5K, electrical shorts, splice resistances, etc. in the course of a superconducting low-field test.
\subsection{Yoke design}
\label{sec:Yoke}
\begin{figure}[h]
\centering
\includegraphics[width=0.95\textwidth]{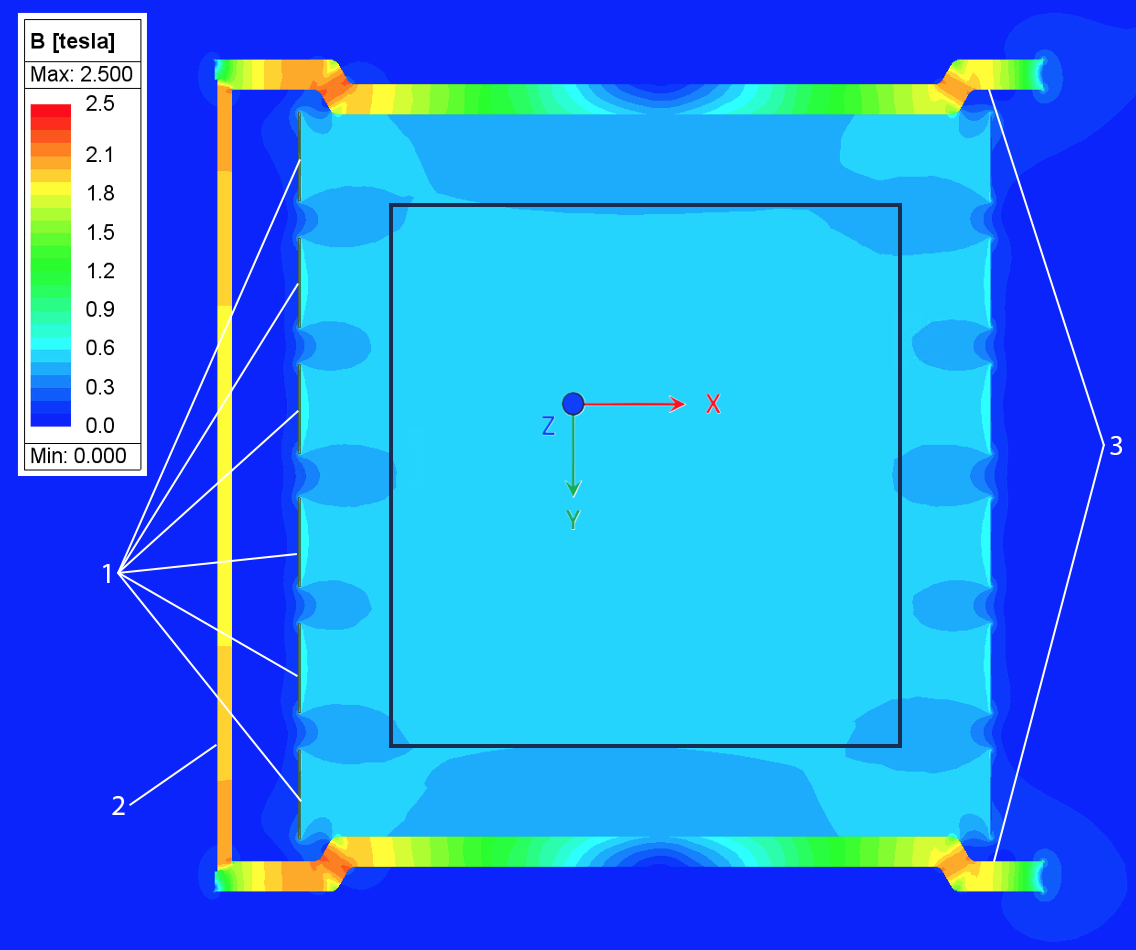}
\caption{Magnetic field strength in ND-GAr: A complete 3D model has been developed and a field map on the horizontal plane crossing the detector center is shown. This section is the most asymmetric, due to the asymmetry of the iron yoke. The TPC volume is defined by the black rectangle. The main components of the magnet are shown: coils (1), iron yoke (2) and stayed heads (3). The neutrino beam travels through the center of the detector, from right to left in this figure.}  
\label{fig:fieldmap}
\end{figure}

%
%As already mentioned, the iron yoke must be asymmetric to guarantee a sufficiently low material budget between ND-LAr and ND-GAr. An entrance window is foreseen on one side of the yoke, making the design of SPY unique. An analogous window on the opposite side has been ruled out due to magnetic containment and to provide the proper amount of material for efficient muon tagging.  A non-magnetic material could have been used, but this would decrease the overall amount of iron in the return yoke thus reducing stray field containment, or would force us to increase the overall weight of the detector, which is already above $1000\unit{t}$. On the side facing the SAND detector, we have considered two alternatives: reducing the asymmetry of the yoke by putting in some non-magnetic material needed for proper muon tagging or, in order to keep the total weight of the apparatus as low as possible, using carbon steel throughout. The proposed solution is a compromise where we use a section of the yoke made of carbon steel, but with reduced thickness. The iron yoke, therefore, will feature an ``entrance window'' facing ND-LAr and a reduced thickness section facing SAND, in an optimized shape to keep the asymmetry small and to minimize the stray field.%
The magnet system for a typical collider detector would have an iron yoke which would include return sections of sufficient cross-sectional area to fully contain the return magnetic field in  iron and thus minimize any fringe fields.  Typically, the return sections would be azimuthally symmetric with respect to the magnetic axis to minimize field distortions.
Fully symmetric return sections are not possible for the SPY magnet because of the two requirements previously mentioned:
\begin{enumerate}
    \item We must eliminate any significant thickness of iron on the upstream face of the yoke to minimize the energy loss of muons passing from ND-LAr to ND-GAr.
    \item A minimum quantity of material is required on the downstream face of the yoke to assist in the discrimination of muons from pions.
\end{enumerate}
The first requirement was satisfied by eliminating the iron from a segment of the front face of the magnet, creating an ``entrance window'' for incoming muons.
A symmetrized magnet design was considered in which the corresponding segment of iron on the down-stream face was removed and replaced by non-magnetic material to meet the muon discrimination requirement while preserving magnetic symmetry.
The remaining iron was then thickened to meet the requirement of field containment. The resulting design failed to meet the space and weight requirements for ND-GAr, however.
The SPY yoke design uses only carbon steel and mirrors the open entrance window on the upstream face of the magnet with a set of thinned return segments on the downstream side.  Because of the required design compromises, detailed simulations were required to validate the final choice of design parameters. A field map within ND-GAr is shown in Figure~\ref{fig:fieldmap}.
%
%
%\clearpage
\section{Stray field analysis}
\label{sec:Stray}
\begin{figure}[h]
\centering
\includegraphics[width=0.98\textwidth]{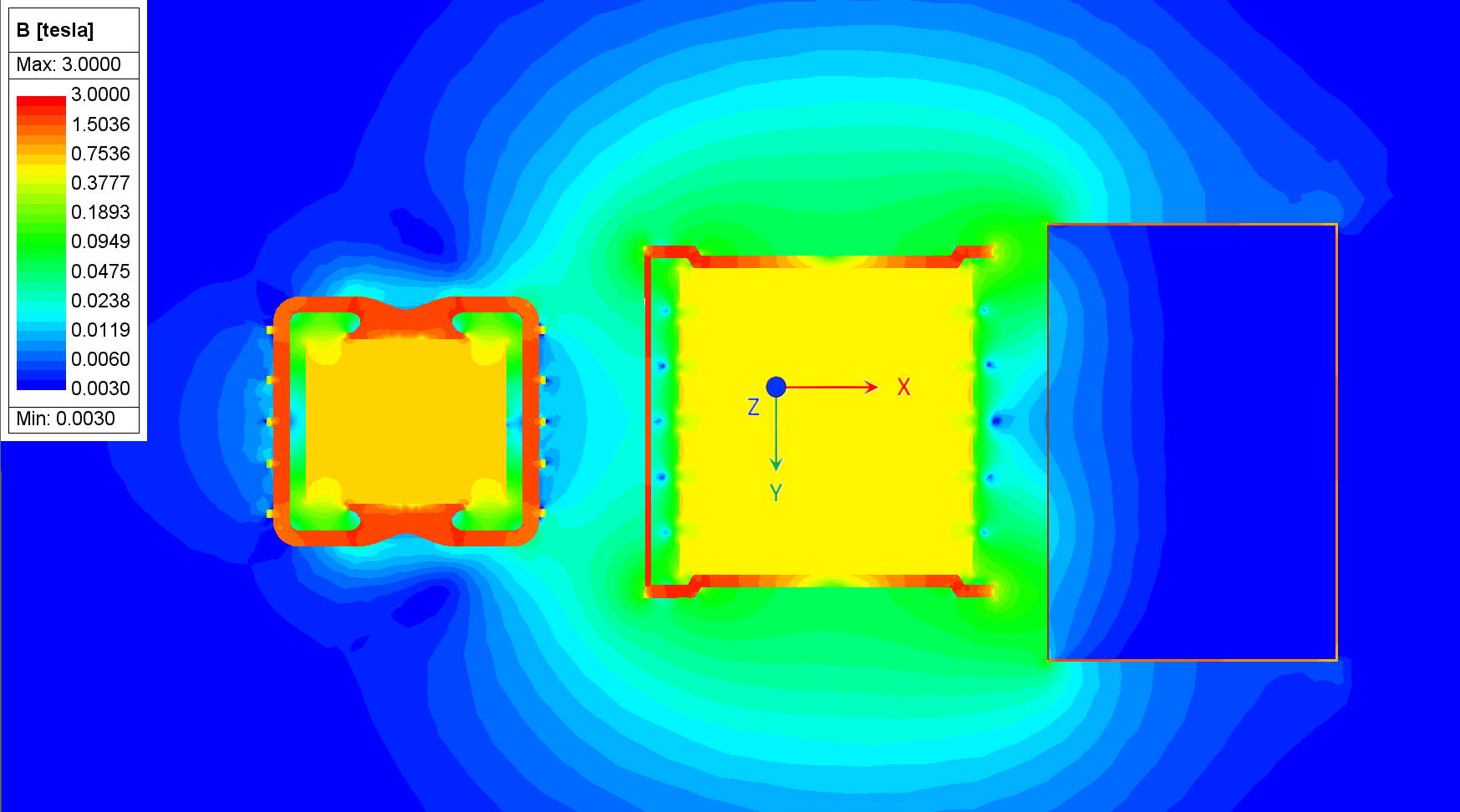}
\caption{Magnetic field strength in the horizontal plane crossing the center of ND-GAr: This model includes the complete set of detectors and several calculations have been performed with different configurations. Here the nominal one is shown, with ND-GAr and SAND magnetic fields at nominal value (and in the same direction) and the three sub-detectors aligned on the beam: SAND, ND-GAr and ND-LAr left to right. For a better visualization of the fringe field, a logaritmic scale between $30\unit{G}$ and $3\unit{T}$ is used. The three sub- detectors have been simulated as follows, from right to left (following the neutrino direction): ND-LAr's carbon steel support structure has been simplified and simulated as a steel box, each face of the same mass of the corresponding support structure part, to simulate the magnetic field shielding; ND-GAr has been simulated in detail with the complete asymmetric iron yoke and the sub-coils (shown with better detail in Figure~\ref{fig:fieldmap}); SAND has been modeled as an iron yoke and coil.  We note that the plane shown here does not cross the center of ND-LAr and SAND due to the inclination of the neutrino beam w.r.t. the horizontal plane and the fact that all the detectors are centered on the beam.}
\label{fig:Field-overall}
\end{figure}

The SPY magnet will operate in close proximity to two other detectors and therefore special attention to stray field is needed. Since the ND-GAr detector will be movable, the cross talk between the three detectors has to be evaluated in different configurations. A field map on the horizontal plane crossing the center of ND-GAr is shown in Figure~\ref{fig:Field-overall}.

\subsection{Field interactions with SAND}
We evaluated the interaction between the SPY magnet system's stray field and SAND. Since SAND is a magnetic spectrometer as well, our analysis must also consider the effect of the SAND field and iron yoke on ND-GAr. The operating parameters for SAND have been obtained from KLOE publications\cite{a_ceccarelli_1997_19360,753226}. 

The magnetic field in SAND is provided by a superconducting solenoid and has a central field design value of $0.6\unit{T}$. Its iron yoke is designed to fully and efficiently contain the stray field.  A small cross talk between the two magnets exists and is due to SPY's stray field interacting with the return iron of SAND.  The contribution from SPY in the active volume inside SAND is negligible  ($\le 0.005\unit{T}$) and is well below SAND's field uniformity specification of 1\%.  The magnetic interaction  with SAND introduces on the order of a $0.001\unit{T}$ variation on the field in SPY with all detectors on axis and the SAND magnet on.  For the various other possible configurations of the Near Detector, i.e. SAND magnet off and ND-GAr on-axis, SAND magnet on and ND-GAr either on-axis or off-axis, we have calculated that the maximum deviation from the field within SPY alone will be less than $0.0025\unit{T}$ in all configurations. This is $0.5\%$ of the design field, and this value is expected only in the peripheral volume of the HPgTPC. It is well within the field uniformity specification (see Table~\ref{tab:fq}).

\subsection{Stray field on ND-LAr}
The stray field on ND-LAr is more critical, due to the small thickness of magnetic material between SPY and the liquid argon TPC.  In the current design, only a few millimeters of carbon steel are in the exit window of ND-LAr.

Due to the complexity of the design of ND-LAr's cryostat support structure, a  simplification of the cryostat had to be introduced in the simulation. The cryostat was modeled as a solid layer of iron of equivalent mass for each side of ND-LAr's cryostat.

\begin{figure}
\centering
\includegraphics[width=0.98\textwidth]{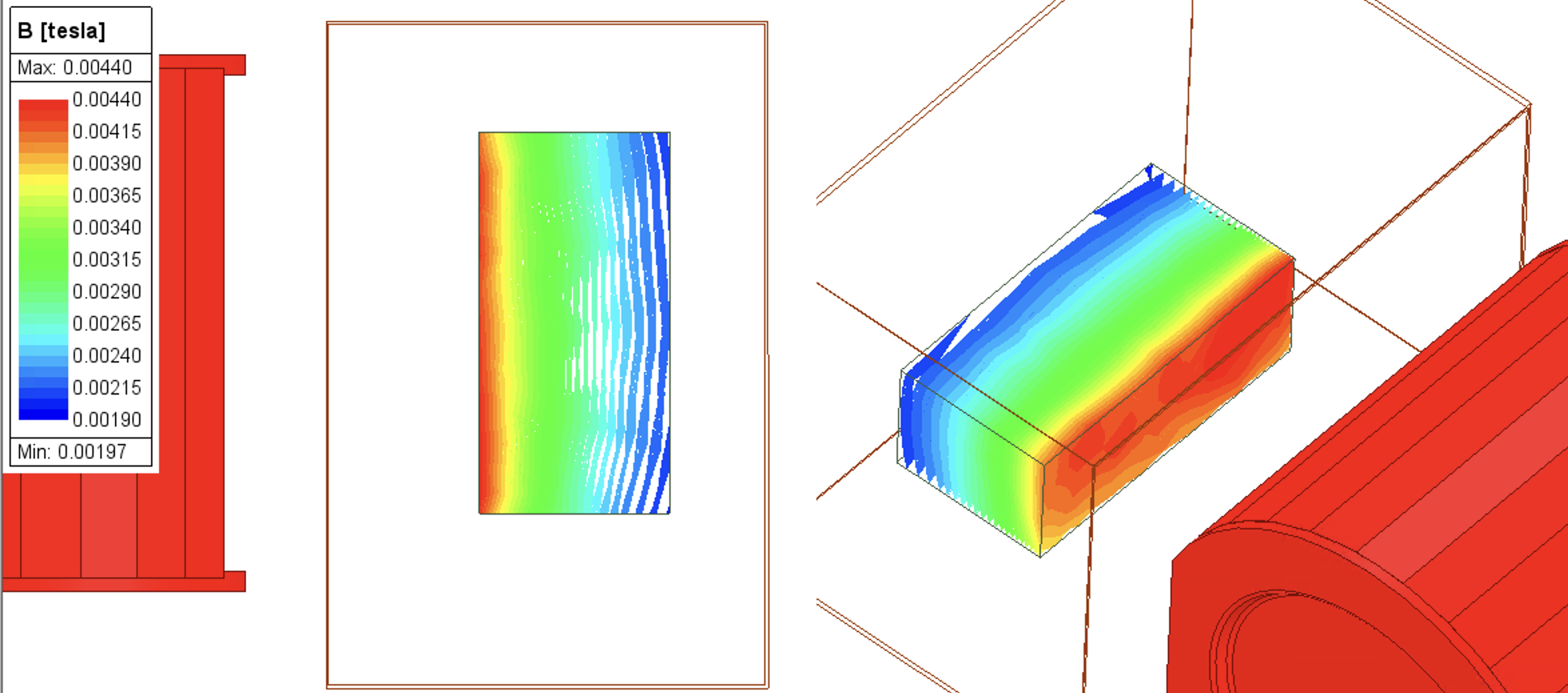}
\caption{The stray magnetic field strength in the fiducial volume of ND-LAr. On the left, the volume is seen from above and the neutrino beam travels from the right to the left of the figure. The support structure for this detector provides a significant complication in the FE model and has therefore been simplified to be just a layer of iron of equivalent mass for each side.  The colored region with field indication represents the 6~m wide by 3.5~m deep fiducial volume of LAr and the outer box in red indicates the the solid iron representation of the cryostat support structure. On the right, the same is shown in isometric view (neutrino beam coming from upper left) to enhance visualization of the three-dimensional behavior of the magnetic field. The calculation indicates that the maximum value of the stray field is 44 Gauss ($4.4\cdot10^{-3}$ T).}
\label{fig:Field-lar}
\end{figure}

The analysis shows that in the current design,  SPY's stray field will produce some field throughout the entire volume of the LArTPC, ranging from $0.001\unit{T}$ to $0.02\unit{T}$.    The field quickly decreases from the side facing ND-GAr to the side from which the neutrino beam is coming. Even in the worst situation, in less than $5\%$ of the active volume does the field exceed $0.01\unit{T}$. In the fiducial volume of the ND-LAr, the stray field is in the range of $0.002\unit{T}$ to $0.005\unit{T}$. A field map in this volume is shown in Figure~\ref{fig:Field-lar}.

\subsection{Stray field on services}
\begin{figure}
\centering
\includegraphics[width=0.95\textwidth]{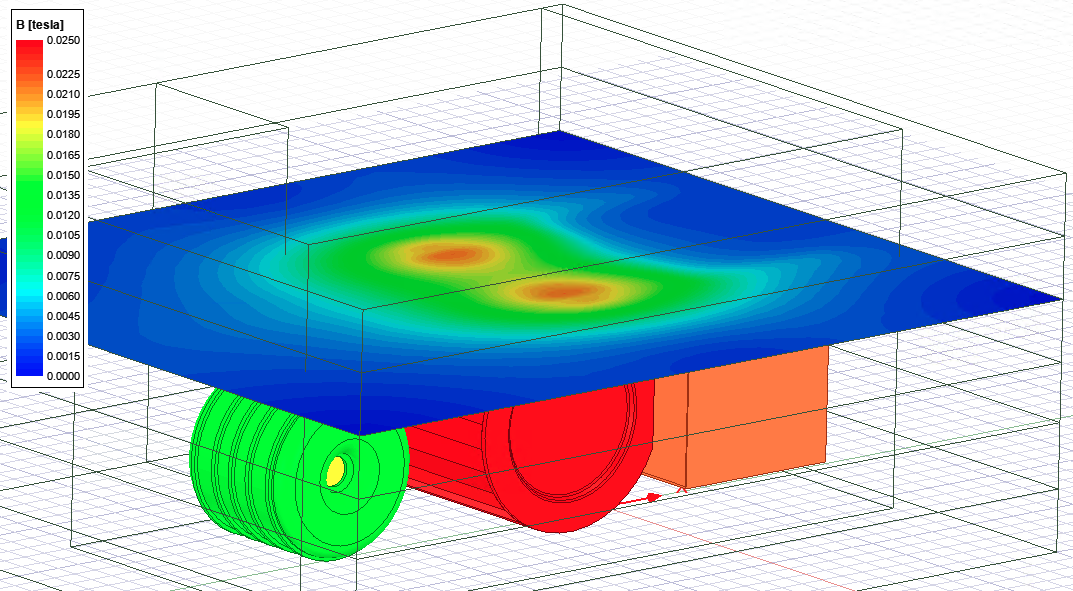}
\caption{Magnetic field strength in a horizontal plane  $5\unit{m}$ above ND-GAr's center is shown. ND-LAr is in orange (first on right) is represented as a carbon steel box with proper thickness to simulate the magnetic properties of the support structure, next is ND-GAr in red and lastly SAND in green. At this distance from the center of ND-GAr, only coolants and gas management material are foreseen (pumps, valves) and the maximum field does not exceed 25 Gauss ($2.5\cdot10^{-3}$ T).}
\label{fig:Field-up10}
\end{figure}

\begin{figure}
\centering
\includegraphics[width=0.95\textwidth]{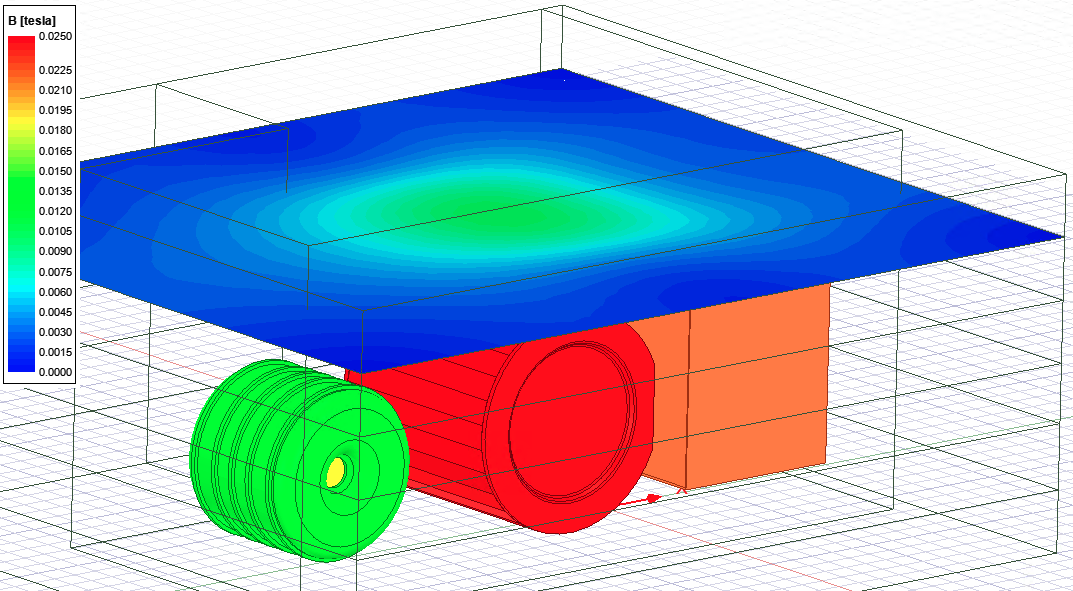}
\caption{Magnetic field in horizontal plane  $7.5\unit{m}$ above ND-GAr's center. Detectors are shown as in figure~\ref{fig:Field-up10}. At this distance from the detector center the presence of front-end electronics is foreseen and the magnetic field does not exceed 15 Gauss ($1.5\cdot10^{-3}$ T).}
\label{fig:Field-up125}
\end{figure}

Several locations for services, including front-end electronics and power transformers, have been evaluated. The most critical volumes are above SAND and above ND-LAr, where  much of the electrical equipment (pumps, sensor electronics, etc.) is expected to be installed. According to our calculations, the stray field in a volume extending $2\unit{m}$ in height above ND-LAr is limited to $0.01\unit{T}$. Two maps of the magnetic field in the whole detector area, on two horizontal planes, are shown in Figure~\ref{fig:Field-up10} and Figure~\ref{fig:Field-up125} at $10\unit{m}$ and $12.5\unit{m}$ height above the detectors' center plane, respectively.

%\clearpage
%
\section{Mechanical design}

\subsection{Design requirements}
The requirements for the mechanical system are as follows: 
%
%\begin{figure}[h]
%\centering
%\includegraphics[width=0.65\textwidth]{Figures/ND-GAr-Section w labels.JPG}
%\caption{General Layout of ND-GAr}
%\label{fig:ND-GArSect}
%\end{figure}
%
%The cryostat that houses the 6-coil solenoid is under vacuum with the coils cooled to $\simeq$ 4.5K. See Figure~\ref{fig:ND-GArSect}.   Initial designs of the pressure vessel utilized a thick-walled cylindrical vessel with spherical heads to accommodate the 10-bar pressure. The vessel heads and associated flanges added excessive length and diameter to the detector, complicating the design of the magnet and the ECAL, and leading to a magnet return yoke that was prohibitively large and heavy.  To address these issues,
%
%\vspace{0.7in}
%
\begin{enumerate}
    \item Provide a vacuum cryostat capable of providing mechanical support and cryogenic environment for the superconducting coils.
    \item The inner wall of the vacuum cryostat must be sufficiently strong to serve as the outer wall of the pressure vessel for the HPgTPC.
    \item The vacuum cryostat walls must be sufficiently strong to provide mechanical support for the ECAL and HPgTPC. 
    \item Provide a carbon steel return yoke for the magnet that can produce a uniform 0.5T central field over the length of the solenoid and contain the fringe fields to the level required by the experiment.
    \item Provide flat carbon steel pole tips for the magnet return yoke that match the magnetic field boundary conditions at the ends of the solenoid and provide the mechanical support for the pressure vessel end flanges.

\end{enumerate}
An additional physics requirement is to measure neutrino interactions in an off-axis position. To meet this requirement, ND-GAr must be able to move perpendicular to the beam.

\subsection{Pressure vessel design analysis approach}
The analysis of ND-GAr's pressurized system was performed to meet the requirements of Fermilab's Environment, Safety, and Health Manual (FESHM), Chapter 5031\cite{FESHM} and the American Society of Mechanical Engineers (ASME) Boiler and Pressure Vessel Code (BPVC). For the ND-GAr system, there are two divisions in the BPVC that could be used to design the vessel, Division 1 and Division 2. Division 1 primarily uses standard design features and manual calculations with a design safety factor of 3.5. Division 2 design methods restricts some design options and requires more inspection to be performed, but allows non-standard designs to be verified with FEA with a design safety factor of 2.4 or 3.0 depending on the vessel class. Due to the unique design and space constraints of the vessel, the design and analysis of ND-GAr's pressurized system has been performed to meet the vessel requirements using the 2019 version of the ASME BPVC, Section VIII, Division 2 \cite{ASME} which will be  referred to as The Code in the rest of this document. To reduce the thickness of components as much as possible to maximize the particle interactions in the detector, the vessel will be designed to the class 2 requirements which uses a design safety factor of 2.4.
%which states:
%
%\begin{itemize}
%    \item Design margin against ultimate tensile strength is 2.4 and
%    \item Design margin against yield strength is 1.5
%\end{itemize}
%
\subsection{Vacuum cryostat}
\begin{figure}[h!]
\centering
\includegraphics[width=0.75\textwidth]{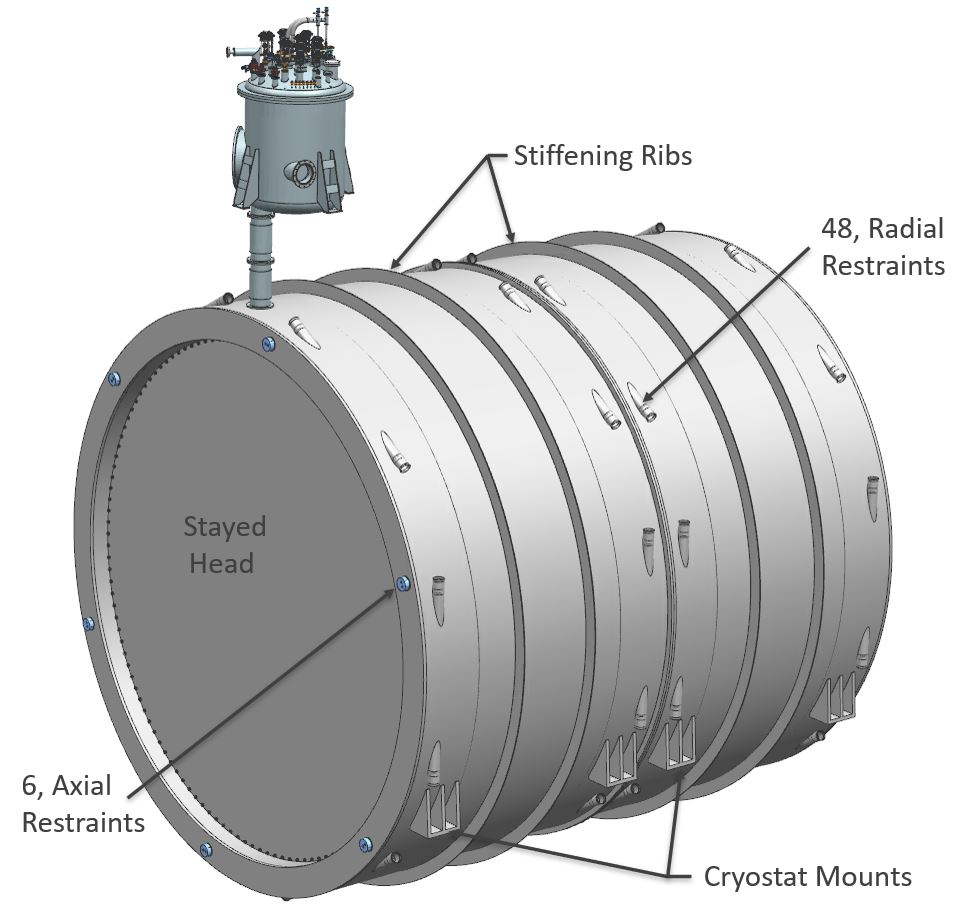}
\caption{Cryostat external features}
\label{fig:Cryostat-w-labels}
\end{figure}

\begin{figure}[h!]
\centering
\includegraphics[width=0.95\textwidth]{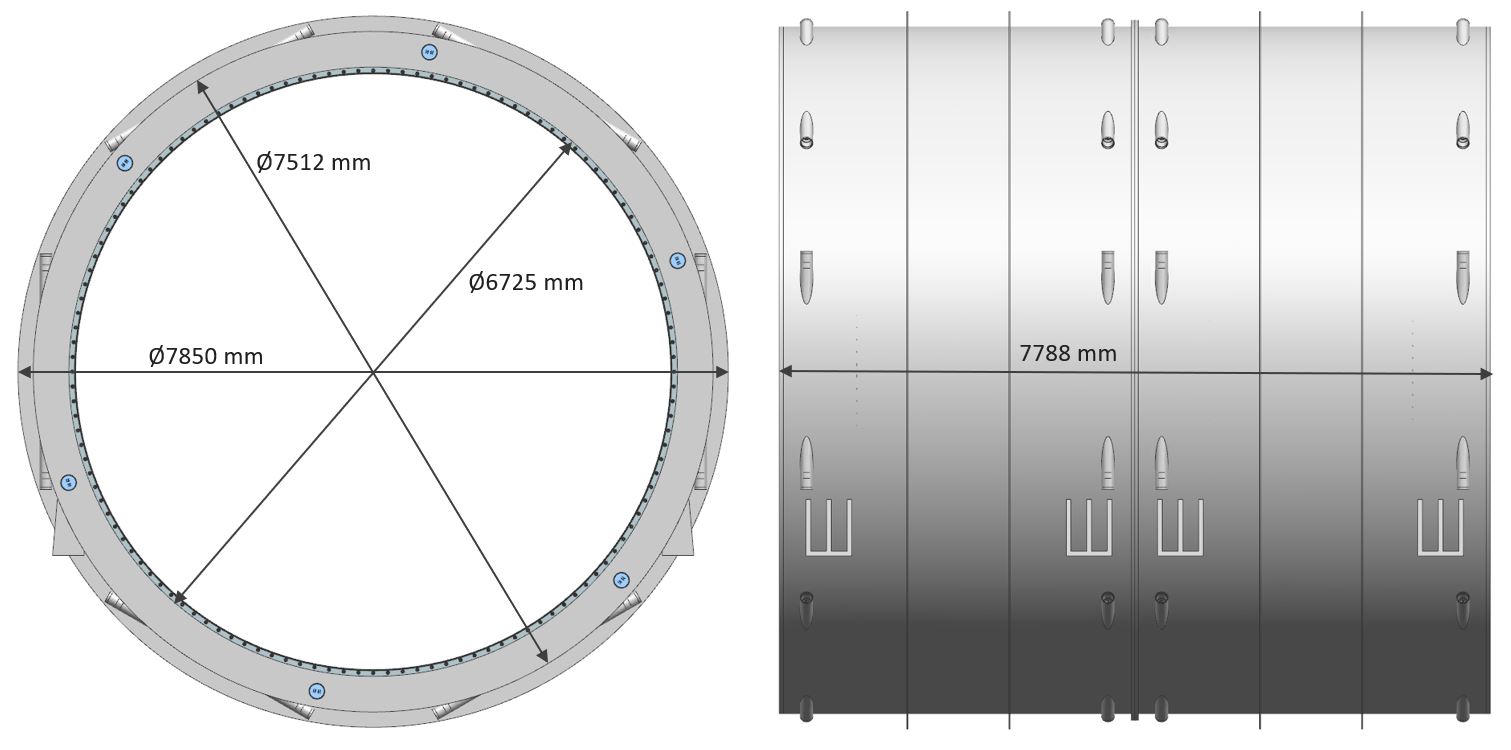}
\caption{Cryostat general dimensions.}
\label{fig:Cryostat-w-dims}
\end{figure}

%The Pressure vessel/Cryostat has evolved from a 2-part design that required field assembly to the current design which requires the magnet to be fully integrated, fabricated and tested at a vendor’s site and then delivered to Fermilab as a single, tested unit.  An integrated magnet design is the preferred choice for cost and reliability considerations, but there are complications with this approach.%
The cryostat design is shown in Figure~\ref{fig:Cryostat-w-labels}. It is 7.512 m in diameter and 7.89 m length with a total weight slightly less than 151 tons.  Figure~\ref{fig:Cryostat-w-labels} also shows the positions of the axial and radial support connections.  See Section~\ref{sec:coldmass} for details on the support rods.
%Because the crane capacity in the Near Detector Hall at Fermilab is limited to 60 tons, a dedicated rigging operation will be required to bring the solenoid into the Detector Hall and position it on its supports.
%Handling the cryostat in the hall will require special rigging as the hall crane capacity is limited to 60 tons.  The design shown here was originally developed under the two-cryostat approach but was modified to accommodate the single-cryostat design.%
%
%Delivering two smaller vessels and associated smaller components keeps the shaft size smaller and less expensive, is easier to transport to Fermilab, keeps each component under the 60-ton limit, but requires assembly and testing at Fermilab in a hall that is not clean enough for this type of assembly work. A cleanroom would need to be erected around the assembly area, adding another expense and complication. The risks to the project are far too high to consider this approach. However, the 
%
The cryostat's overall dimensions are shown in Figure~\ref{fig:Cryostat-w-dims}.  Due to the span of the vessel, stiffening ribs are used to strengthen the outer shell. The stiffening ribs have a thickness of 12.7 mm and have an outer diameter of 7.85 m. The stayed heads and cryogenic feed-can are not included during the initial installation of the cryostat.

\begin{figure}[h!]
\centering
\includegraphics[width=0.95\textwidth]{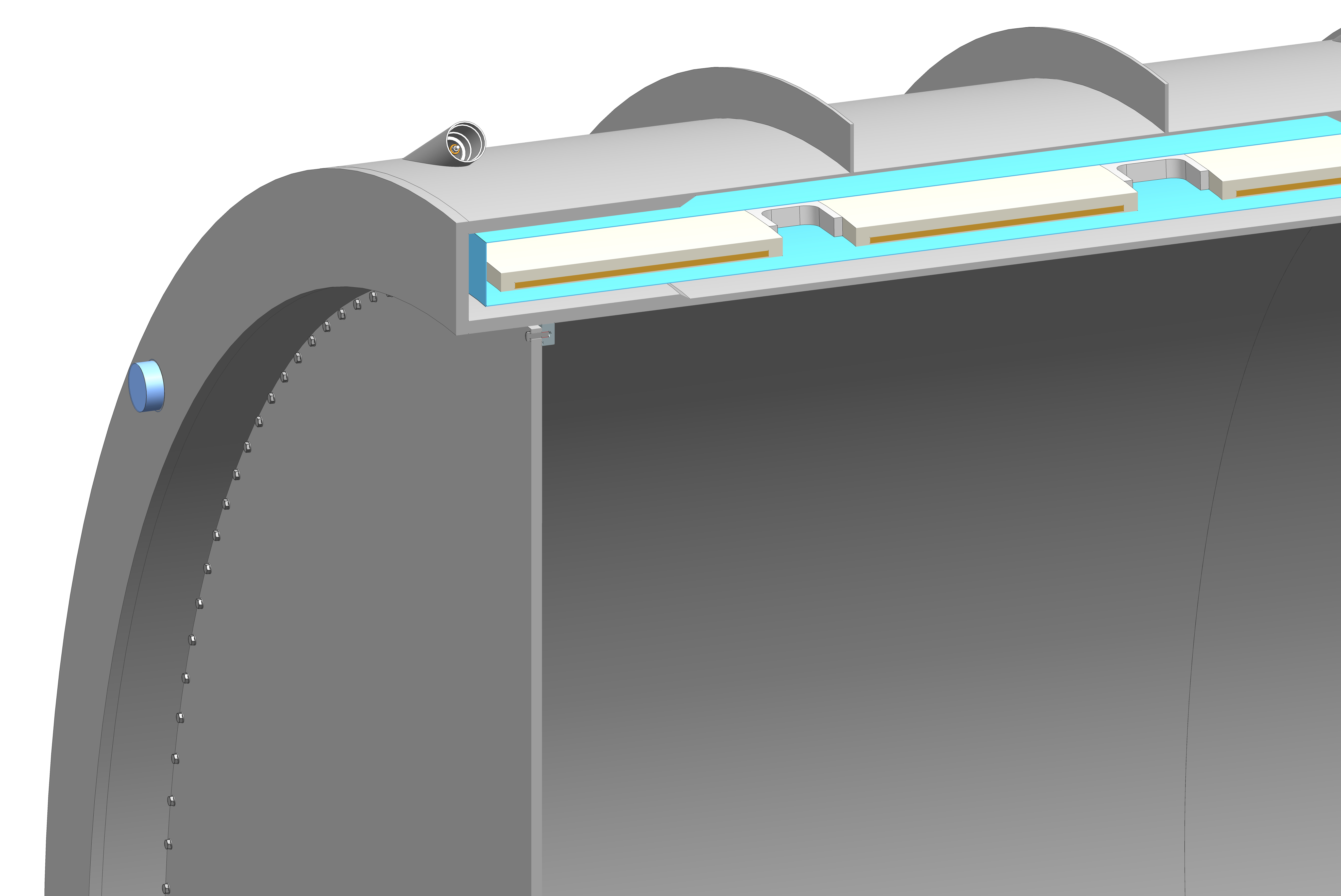}
\caption{Sectioned cryostat blowup showing three of the six magnet coils and the heat shield (in turquoise).}
\label{fig:Cryostat-w-section}
\end{figure}

The cryostat is designed to serve as an insulated vacuum vessel that houses the six internal superconducting coils and the radiation shield, see Figure~\ref{fig:Cryostat-w-section}.  It also must support the neutrino detector in its bore which operates in a 10-bar atmosphere.  Note: See section~\ref{sec:coldmass}, Figure~\ref{fig:Coil_dims} for more details on the coldmass. The inner shell of the cryostat must accommodate this 10-bar pressure. The 38.1 mm thick (1.5 inch) flat heads at each end of the cryostat cannot withstand the 10-bar pressure on their own. The design requires that the heads be supported by the yoke end plates using 798 stays per head.  See Section~\ref{subsec:Yoke}. 

We expect that the solenoid and its vacuum cryostat in SPY will be fabricated, assembled, and tested at the vendor site and delivered to Fermilab as a working unit. 
Although consideration was given to a design based on delivering smaller sub-assemblies to Fermilab and completing the final assembly underground, after considering cost, reliability, and logistical complications, the vendor-integrated assembly emerged as the preferred option.

%The cryostat is a welded and bolted assembly fabricated from 316LN stainless steel and was designed using the ASME Boiler and Pressure Vessel code, Section VIII, Division 2, Design by Analysis, Elastic-Plastic analysis method, see section 4.4, ANSYS analysis. The cryostat is mounted in the yoke at 8 locations, but this design parameter can also be readdressed with a single cryostat design approach – potentially using only six supports.
%
%
% --Not sure this line should appear at this location...
%The stayed heads add a complication to the cryostat design. Due to the large flat surface area, there is a large force and bending moment at the inner mounting flange that creates a potential failure mode. Refer to Section~\ref{sec:FM} for more details. %
%\clearpage
%
%
\subsubsection{Vacuum failure analysis}
Since the magnet's cryostat also provides pressure containment, we did a preliminary analysis regarding an insulating vacuum failure \cite{Voirin}.  A loss of insulating vacuum will allow a large amount of heat to be transferred among the cold coil assembly, the liquid nitrogen shield, and the room temperature vacuum cryostat. This heat transfer would cause thermal shrinkage of the cryostat, and possible leakage at the large flanges which connect the stayed heads to the magnet cryostat providing pressure containment. This could result in an oxygen deficiency hazard and, depending on the rate of loss, cause damage to the HPgTPC. This analysis is used to quantify the resulting temperature changes of the cryostat versus time in the event of a loss of insulating vacuum.  
Simple energy-balance calculations show an equilibrium temperature of 220K. 

Preliminary structural FEA modeling using uniform thermal shrinkage shows that the proposed flexible bolted connection on the stayed-head flange is capable of remaining sealed.  
%In reality, the temperature profile will be non-uniform, and there are other structures where the real resulting temperature in the loss of vacuum scenario would be needed for structural analyses.
We used commercial Computational Fluid Dynamics code to simulate the partially laminar, partially turbulent, transient buoyancy driven convection of the incoming air, and the heat transfer among the components and from the cryostat to the room temperature yoke. The analysis shows that the cryostat falls in temperature very slowly, taking approximately 2000 minutes to reach its minimum value. The average temperature of the cryostat at 2000 minutes is 271.5K, a drop of $\simeq$ 21.5K from its original 293K Temperature.  The minimum temperature of the cryostat was 260 K, which was located at the bottom of the inner gaseous argon surface, near the center of the cryostat.  We note that although a design of HPgTPC gas system is not currently available, we believe that in an emergency scenario, controlled venting to 1 bar from 10 bar can be accomplished in a time short compared to 2000 minutes.

\begin{figure}[h]
\centering
\includegraphics[width=0.95\textwidth]{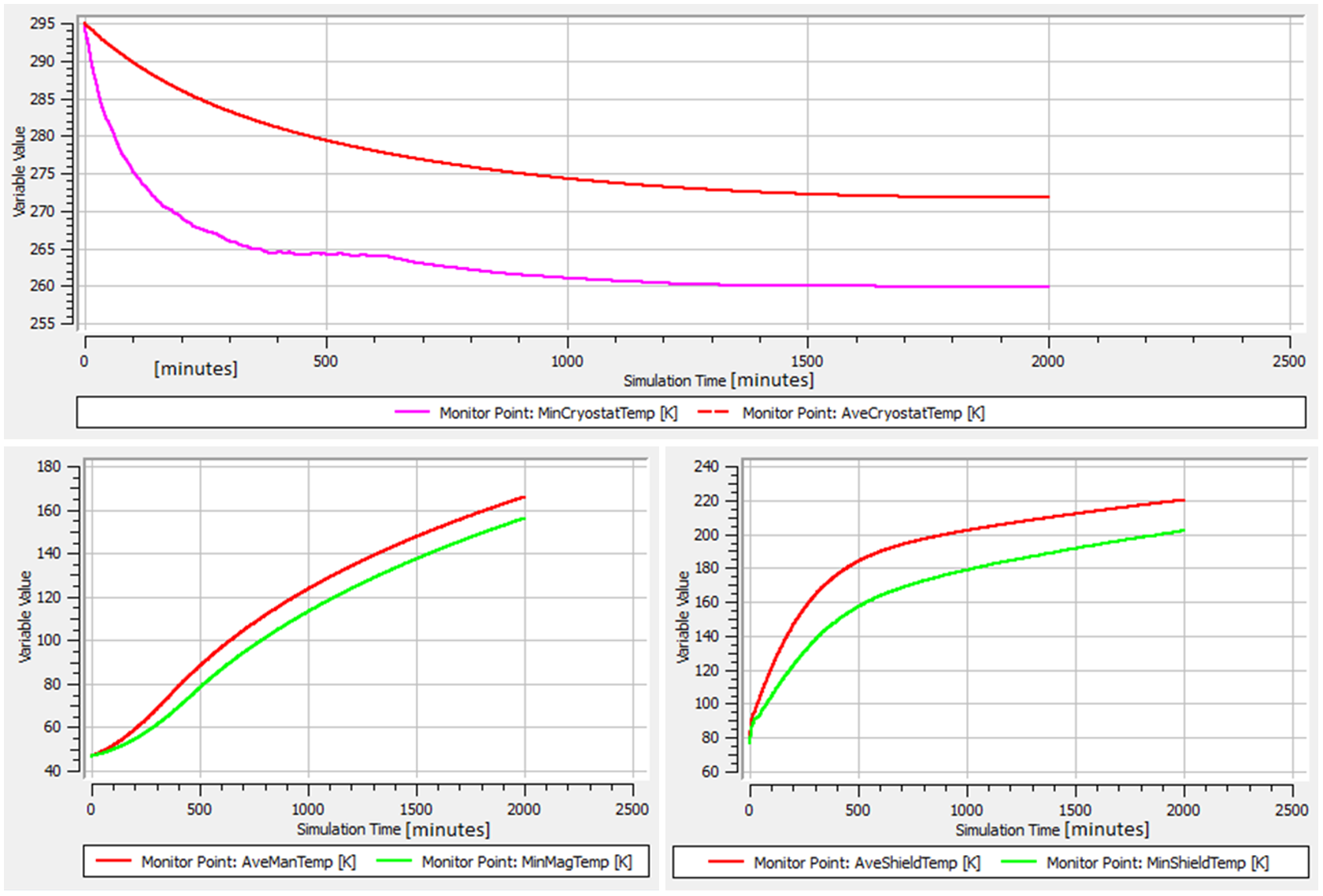}
\caption{Transient average and minimum temperatures of the cryostat (Top), magnet (Bottom, left), and LN2 shield (Bottom, right)}
\label{fig:Thermal4}
\end{figure}

\begin{figure}[h]
\centering
\includegraphics[width=0.95\textwidth]{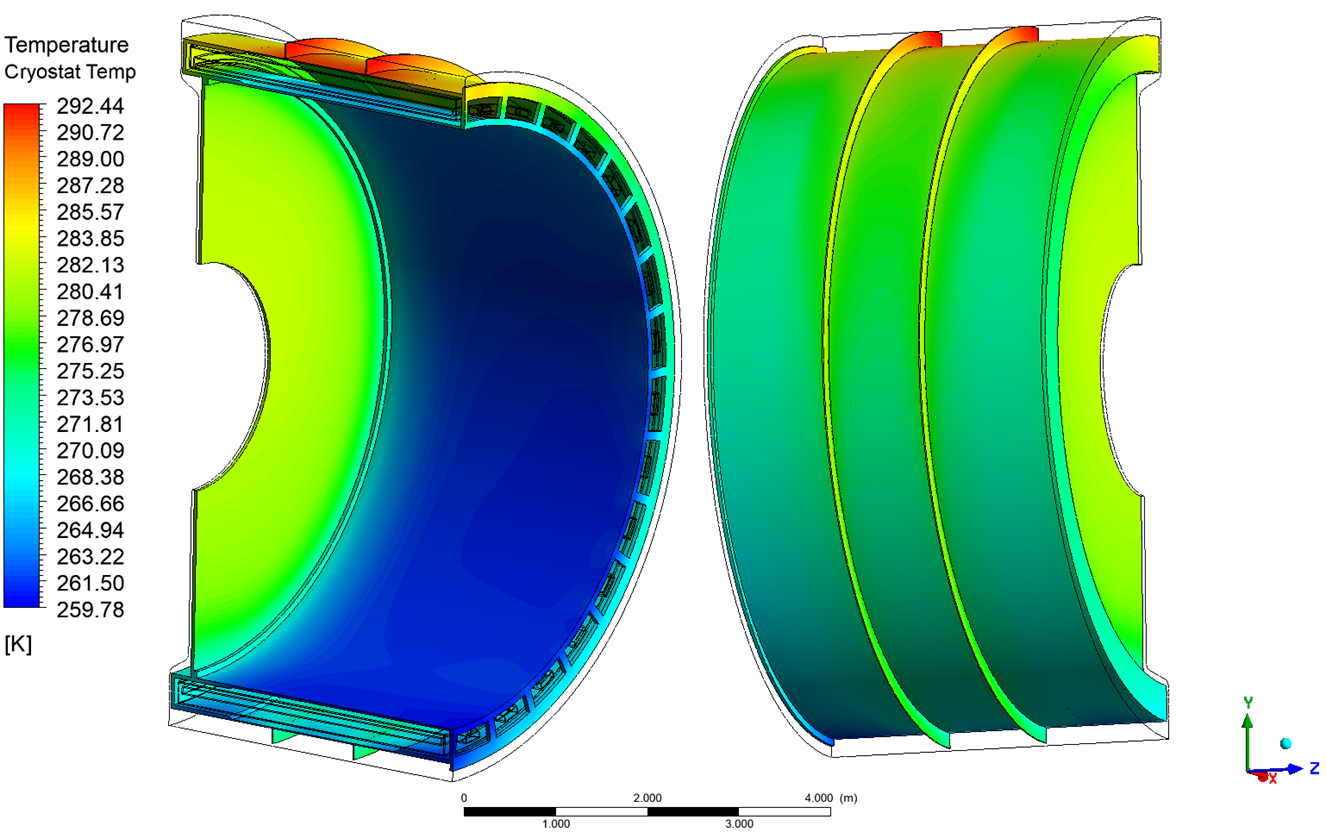}
\caption{Temperature profile of the cryostat at the end of the 2000 minute simulation when minimum temperature had been reached.}
\label{fig:Thermal5}
\end{figure}

\begin{figure}[h]
\centering
\includegraphics[width=0.95\textwidth]{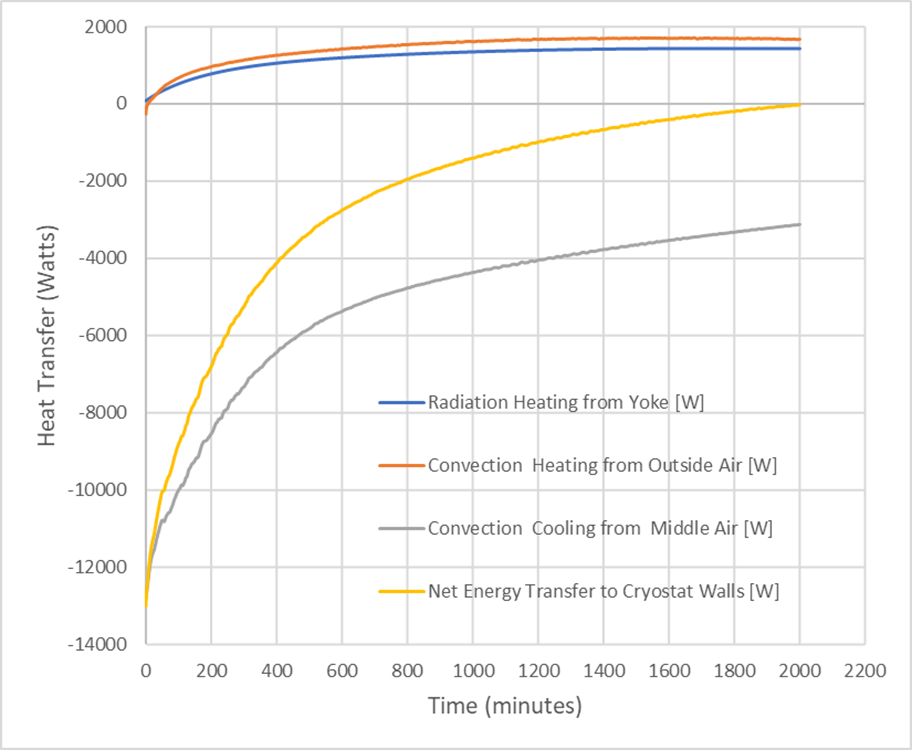}
\caption{Heat transfer to the cryostat from the 3 sources described in the text. In addition, the total net energy transfer to the cryostat (sum of the three) is also shown.}
\label{fig:Thermal7}
\end{figure}

Figure~\ref{fig:Thermal4} shows the transient average and minimum temperatures of the cryostat, the LN$_2$ shield, and the coils in degree K.  The temperature profile of the cryostat at the end of the 2000 minute simulation is shown in Figure~\ref{fig:Thermal5}.
Figure~\ref{fig:Thermal7} shows the heat transfer to/from the cryostat from three sources.  These are: radiation heating from the yoke, convective heating from the outside air, and convective cooling. The convective cooling is via the internal air between the cryostat, the LN$_2$ shield and crossover part of the magnet.  The net energy transfer to the cryostat is the sum of the three.
%
%\clearpage
%
\subsection{Coldmass}
\label{sec:coldmass}
\begin{figure}[h!]
\centering
\includegraphics[width=0.95\textwidth]{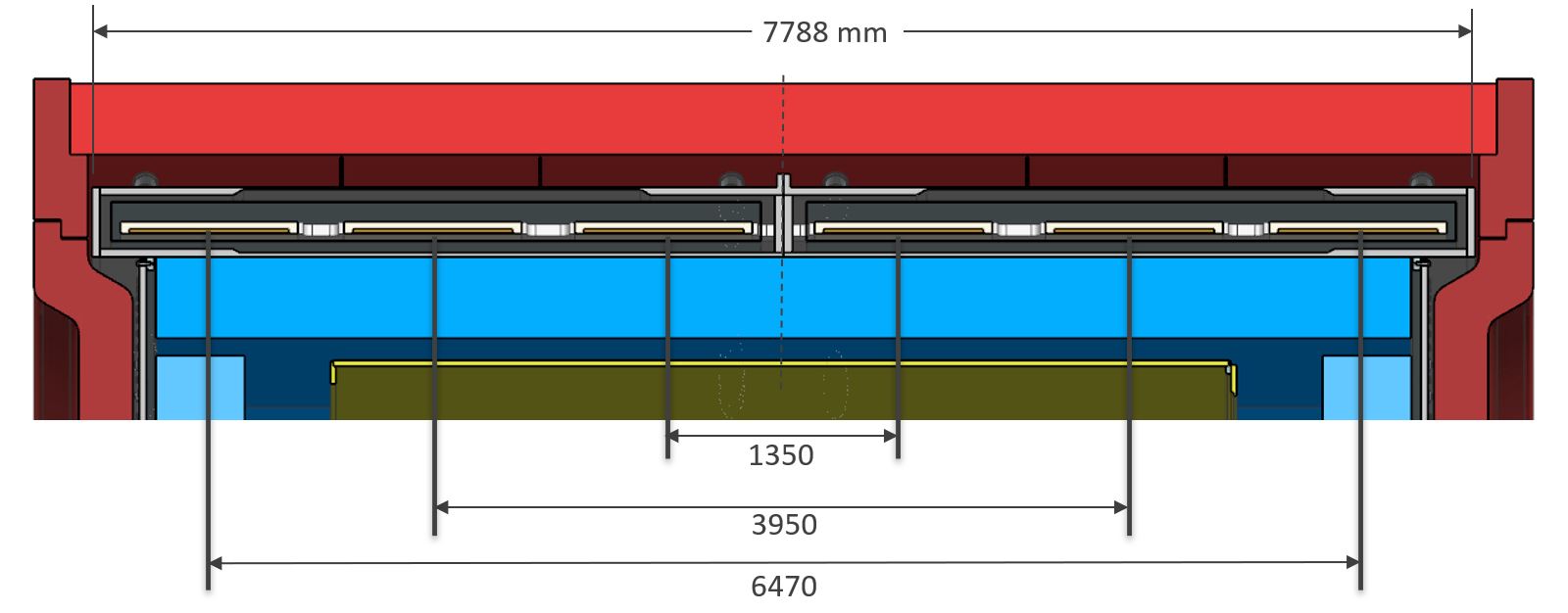}
\caption{Coil layout and dimensions}
\label{fig:Coil_dims}
\end{figure}

 The coldmass consists of six superconducting coils surrounded by a 4 mm thick aluminum thermal shield as outlined in Section~\ref{sec:Coil}. The coil and bobbin assembly weighs ~30 tons and is supported inside the cryostat both radially and axially.  The assembly's outer diameter is 7040 mm with an inner diameter of 7000 mm. The coil layout is shown in Figure~\ref{fig:Coil_dims}. 
 
\begin{figure}[h!]
\centering
\includegraphics[width=0.75\textwidth]{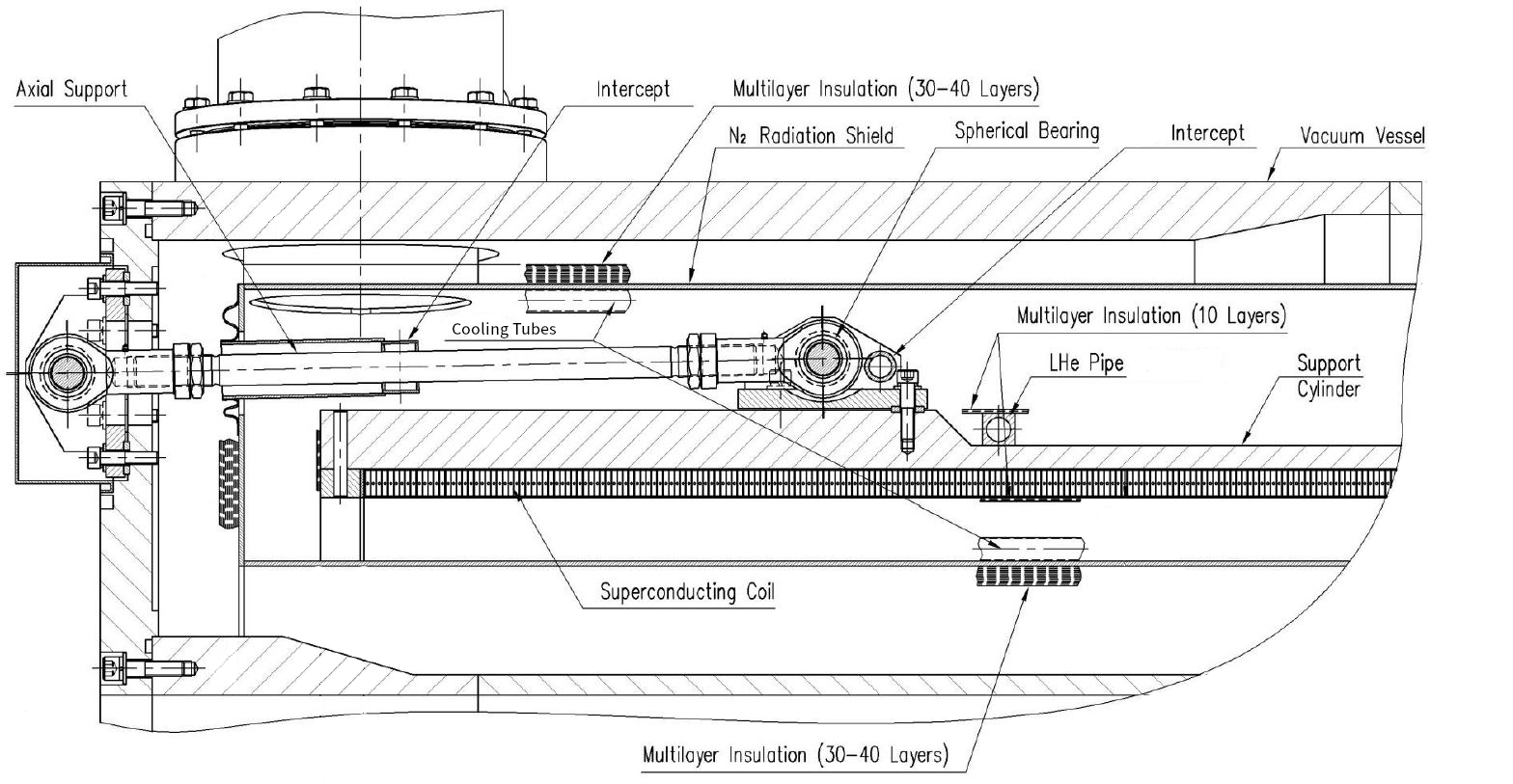}
\caption{Axial Support Rods}
\label{fig:Axial_Supports}
\end{figure}

 The coils are connected in series, and due to the nature of the symmetry, a force balance in the magnet is achieved. Having the coils powered in series also means that potential coil failures will force the power to ramp down uniformly. Any potential imbalance of force produced by the proximity of the SAND magnet is constrained by six axial restraints mounted at only one end of the magnet.  We do not have a detailed design for these supports in SPY, but the configuration used in the JINR MPD solenoid~\cite{MPD-JINR} as shown in Figure~\ref{fig:Axial_Supports} is appropriate for SPY also. 

\begin{figure}[t!]
\centering
\includegraphics[width=0.96\textwidth]{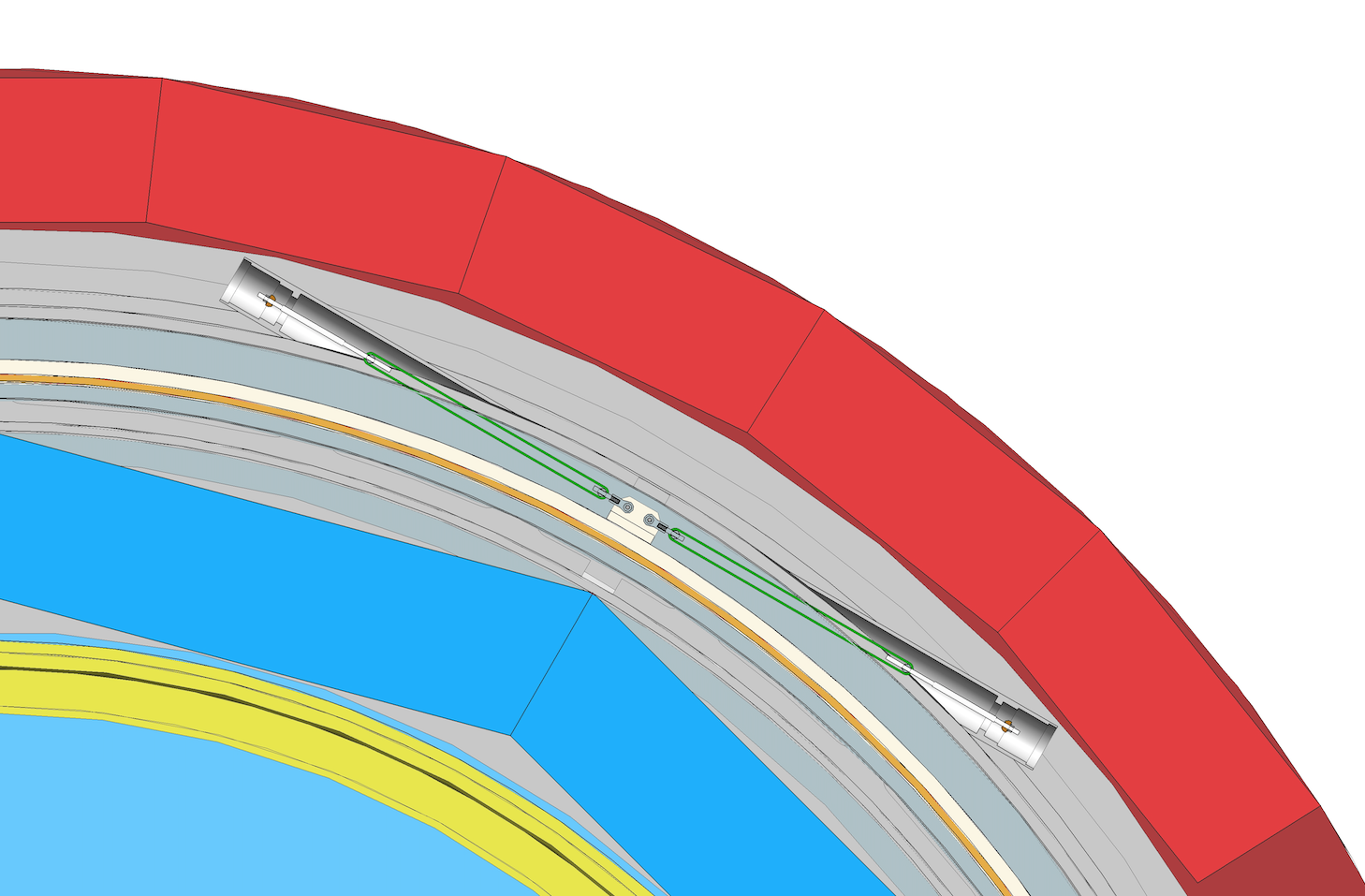}
\caption{Radial support configuration}
\label{fig:Coil-Restraints-Radial}
\end{figure}

The radial supports, as shown in Figure~\ref{fig:Coil-Restraints-Radial}, are designed to support the
coil assembly’s dead load, the magnetic load created from the proximity of the SAND magnet, and the loading due to the non-symmetric yoke design. In addition, forces develop on the radial supports when the magnet shrinks due to cooling from room temperature to operating temperature. To minimize and potentially completely cancel out the loading due to shrinkage, the radial supports are designed at an angle and with ball-joint end connections which allows the supports to rotate as the coil bobbin shrinks radially inward. With this design, the support rods will simply rotate to the new alignment position and not develop any additional axial loading.

\begin{figure}[h!]
\centering
\includegraphics[width=0.95\textwidth]{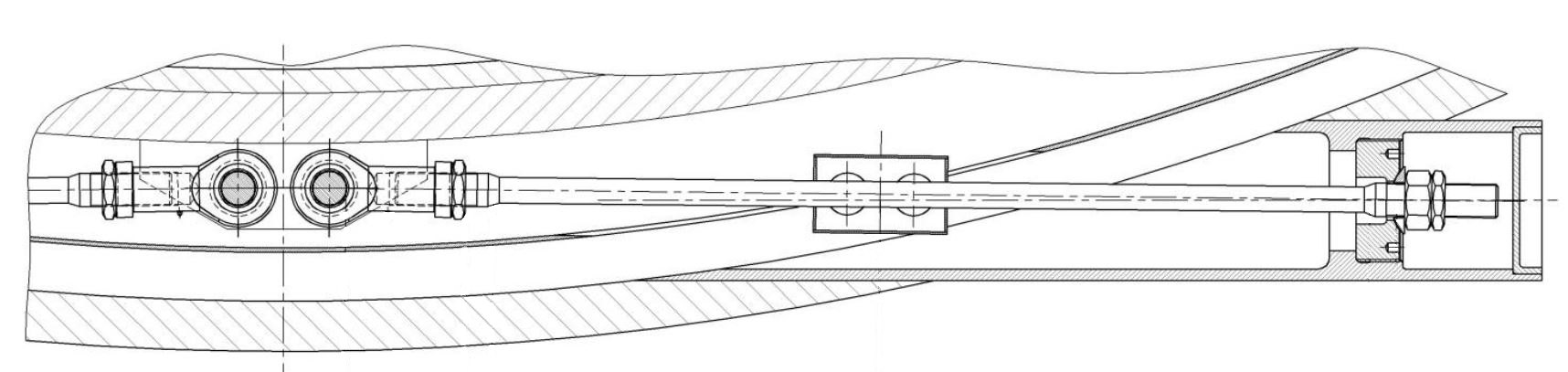}
\caption{Radial support - Option 1}
\label{fig:Radial-Rod-Assy1}
\end{figure}

\begin{figure}[h!]
\centering
\includegraphics[width=0.95\textwidth]{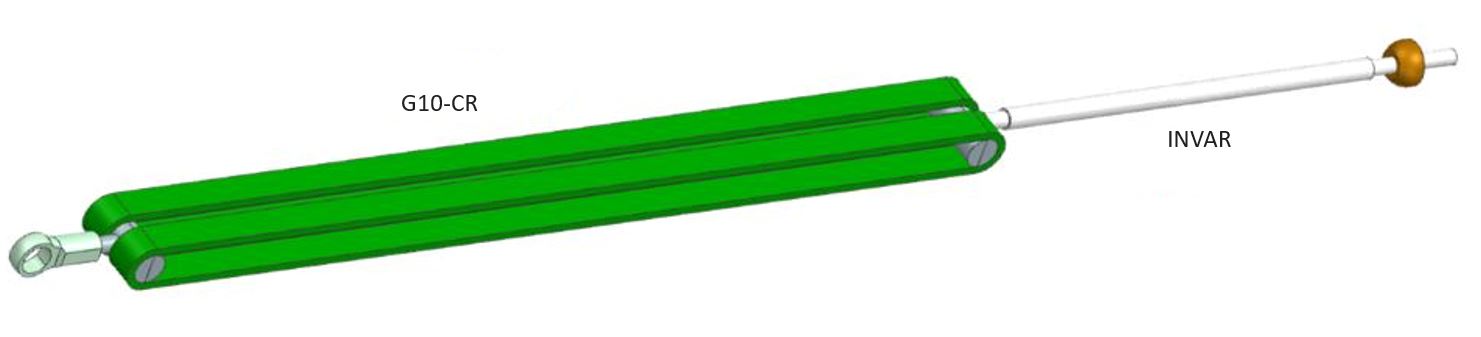}
\caption{Radial support - Option 2}
\label{fig:Radial-Rod-Assy}
\end{figure}

We have developed two viable options for the radial supports. Both options utilize an intermediate heat sink operating at between 50K and 80K. Both designs also support the dead load of the coils with the downward hanging vertical supports. The radial supports  maintain the circularity and center the coil bobbin assembly.  They will also withstand the anticipated magnetic forces on the assembly.
In design option 1 (Figure~\ref{fig:Radial-Rod-Assy1}), solid invar rods are used with a thermal sink as indicated above. 
%as shown in Figure~\ref{fig:Radial Rod-Assy}. 
This design option is simple to design, analyze, and manufacture but produces a large heat leak.  Design option 2 for the radial supports is an assembly as shown in Figure~\ref{fig:Radial-Rod-Assy}. The support has two components which make the transition at the 50K to 70K thermal sink.  This allows for very efficient heat shunting. One part of the support is constructed from an invar (or a similar material) rod, while the other part is constructed from G10 or carbon fiber thermal straps. The lengths
of the stages can be fine tuned by adjusting the mounting angle to optimize the strength of the support and to minimize the stresses. Commercial sources for this type of strap exist. This option greatly reduces the thermal leak and allows for some assembly adjustment during construction and maintenance. 

%
%\begin{figure}[h!]
%\centering
%\includegraphics[width=0.75\textwidth]{Figures/Radial Rod Assy 2.JPG}
%\caption{Radial support rod - Option 1}
%\label{fig:Radial Rod-Assy}
%\end{figure}
%
%
Regardless of the axial and radial supports used in the final design, attention must be given to shipping and installation requirements. Either of these support options must withstand shipping loads or additional shipping restraints will need to be incorporated into the design that can be removed after installation.

%\subsection{Pressure containment}

%The pressure volume of the cryostat is separated into three regions. The three regions can be seen in 

%%
%\begin{figure}[h!]
%\centering
%\includegraphics[width=0.90\textwidth]{Figures/3-detector_2-24.png}
%\caption{SPY Magnet System Pressure Regions}
%\label{fig:ND}
%\end{figure}
%%

\subsection{Finite element analyses (FEA) }
\label{sec:FEAAnl}
To determine the safety of the system, a combination of design by rule and elastic-plastic design by analysis methods were used. Elastic-plastic analysis methods examines the plastic capacity of the model by factoring in safety factors stored inside the loading factors($\beta$) and load combinations. Analysis was performed using the load factors of a Class 2 vessel, $\beta$ = 2.4, and were derated by 0.85 to account for the joint efficiencies of the welds. To meet the FESHM 5031 requirements, an additional derating factor of 0.8 was applied to the loads. 
% 
%Analysis of the ND-GAr pressurized system was performed to meet the requirement's of the FESHM 5031 \cite{FESHM} and the BPVC VIII, Division 2 \cite{ASME} codes. Analysis was performed using the safety factors of a Class 2 vessel and derated by 0.85 to account for the joint efficiencies of the welds. To meet the FESHM 5031 requirements, the safety factor was further derated by 0.8. 
%
When it was possible, initial calculations were performed using part 4 of the ASME VIII Div.~2 specifications and the calculations were verified using the Design by Analysis Methods as described in part 5 of the ASME VIII Div.~2 specifications. Due to the complex loading conditions and asymmetrical design of the cryostat, the elastic-plastic stress analysis process was performed for all components as recommended in 5.2.1.2 of The Code. By using an elastic-plastic material model, limits are set based on the allowable plastic strain the assembly can withstand instead of an allowable stress limit. 

There are generally four steps that are needed to be performed to show an acceptable design:
\begin{enumerate}
    \item Protection against plastic collapse
    \item Protection against local failure
    \item Protection against collapse from buckling
    \item Protection against failure from cyclic loading
\end{enumerate}
Two general areas of analysis were performed: an analysis of the cryostat head and an analysis of the shell thicknesses of the cryostat.  Protection against plastic collapse, local failure, and collapse from buckling have been considered in this initial design. 

To show protection against plastic collapse, the loads that are applied on the model are scaled by the loading factor $\beta$.  If the model is able to converge on a solution it is shown to meet the plastic collapse requirements. As scaled loads are used to determine the acceptable limit of the vessel, the resulting deformation shown will be higher than what will occur in the actual components. To show protection against local failure the model is solved at $\beta$ = 1.7. Additional load deraiting is performed using the FESHM and weld joint efficiencies. Solving the analysis model with loads scaled to the local failure load factor, the limiting triaxial strain can be found and compared to the equivalent plastic strain in the model. Modifying the equations in chapter 5.3.3 of The Code, protection against local failure is shown when the following expression is satisfied at every point in the model:  

\begin{equation}\label{modified}
     \epsilon_{limit} = \frac{\epsilon_{peq}}{\epsilon_{Lu}*exp[-(\frac{\alpha_{sl}}{1+m_2})(\frac{\sigma_1+\sigma_2+\sigma_3}{3\sigma_e}-\frac{1}{3})]} \leq 1
\end{equation}
where:
\begin{itemize}
    \item $\epsilon_{limit}$ is the local strain limit.
    \item $\epsilon_{peq}$ and $\epsilon_{Lu}$ are the equivalent plastic strain and limiting triaxial strain.
    \item $\sigma_{1}$, $\sigma_{2}$, and $\sigma_{3}$ are principal stresses.
    \item $\sigma_{e}$ is the equivalent stress.
    \item $\alpha_{sl}$ and $m_2$ are material dependent properties.
\end{itemize}

Protection against buckling was determined using Type 3 methods as per 5.4.1.2 of The Code. In this assessment, an eigenvalue buckling model is solved to determine imperfections in the model that could create buckling conditions. The imperfections are then scaled based on the tolerances of the assembly to create a new 3D model of the assembly. The new 3D model is then re-examined to the plastic collapse requirements. If the model is able to converge on a solution with the exaggerated imperfections, the model meets the protection against buckling requirements.

%Each step was examined when performing the analysis of the ND-GAr System, however, minimal attention was paid to the protection against failure from cyclic loading as it relates to the lifetime of the assembly. As there are many smaller details that have not been defined yet, that will determine the overall lifetime of the assembly.  This should be evaluated when the details are better defined. 

%There were two general areas of analysis that were performed: The analysis of the cryostat head and the analysis of the shell thicknesses of the cryostat.
%
\subsubsection{Pressure vessel head analysis}
\label{sec:PVHA}
\begin{figure}[h!]
\centering
\includegraphics[width=0.80\textwidth]{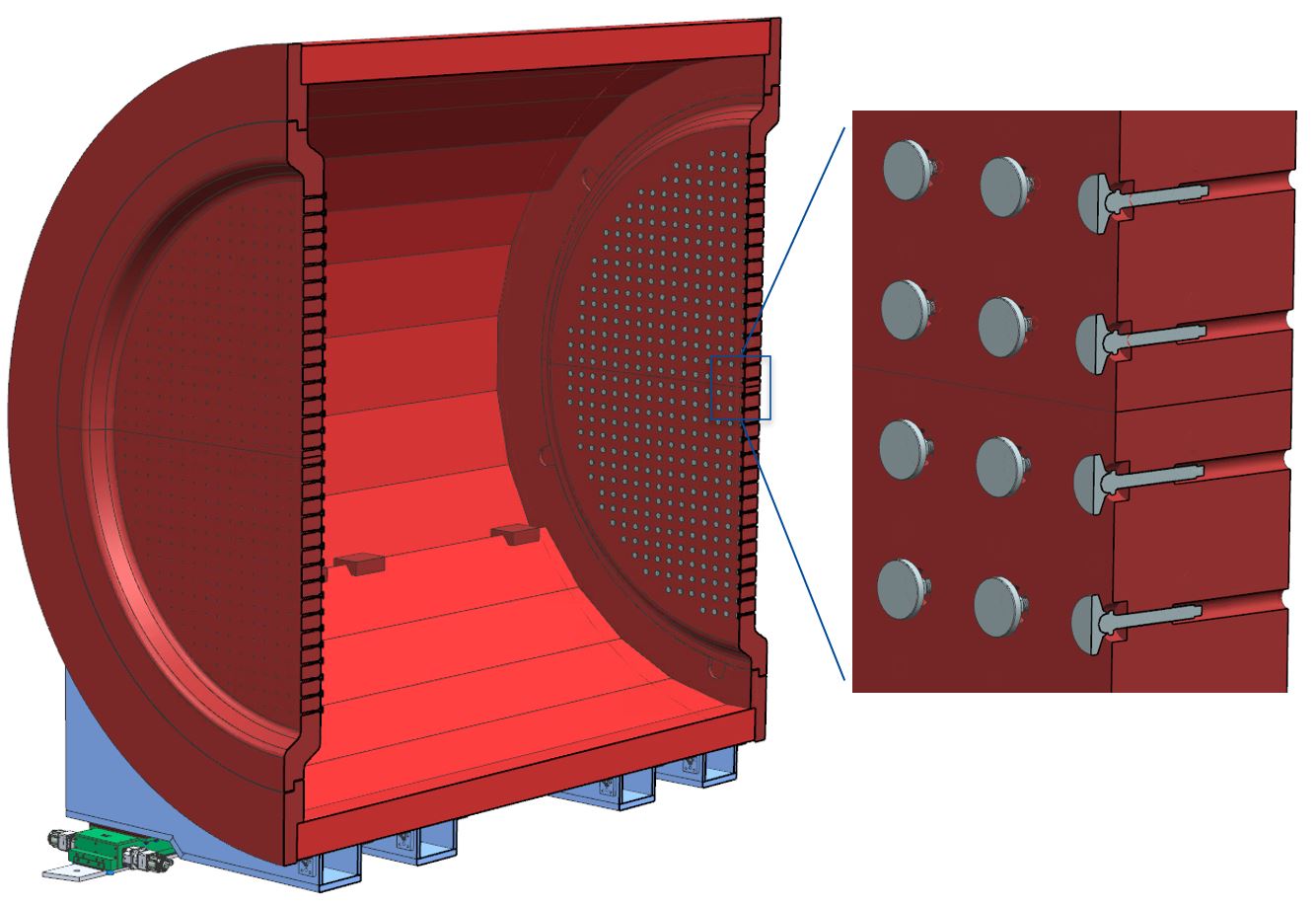}
\caption{Stayed head conceptual design.}
\label{fig:Stayed}
\end{figure}

 %Performing some initial calculations on the required thickness of the head indicates that to resist the 10-bar operating pressure, the head would be required to be approximately 11 inches thick. This is not an acceptable thickness due to the amount of space available, the operating parameters of the ND-GAr system, and the downstream design consequences of using a large head. %
The stayed head design for the integral pressure vessel of SPY is based on the observation that the 0.28 m thickness required for a flat pressure vessel head is comparable to the thickness of carbon steel needed for the pole tips of the return yoke.
To reduce the required thickness of the pressure vessel heads, a grid of 798 3/4" stay bolts spaced on 7 inch centers through each pole tip will be used to brace the pressure vessel heads against the magnet yoke, allowing the heads to be relatively thin. The stay bolts are simple threaded rod leveling pads. See Figure~\ref{fig:Stayed}.
Following the rules for stayed heads, a range of parameters was defined. Using a fixed constraint at each of the stay bolt locations, the resulting minimum spacing pitch distance with respect to the minimum plate thickness was determined.
Following the results of the calculations, a simplified model of the stayed flathead was created using the derived parameters. Stay bolt spacing and sizing was examined in the model to conceptually verify the design. Further refinement and optimization is needed to finalize the design. 

\begin{figure}
     \centering
     \begin{subfigure}[b]{.4\textwidth}
         \centering
         \includegraphics[width=\textwidth]{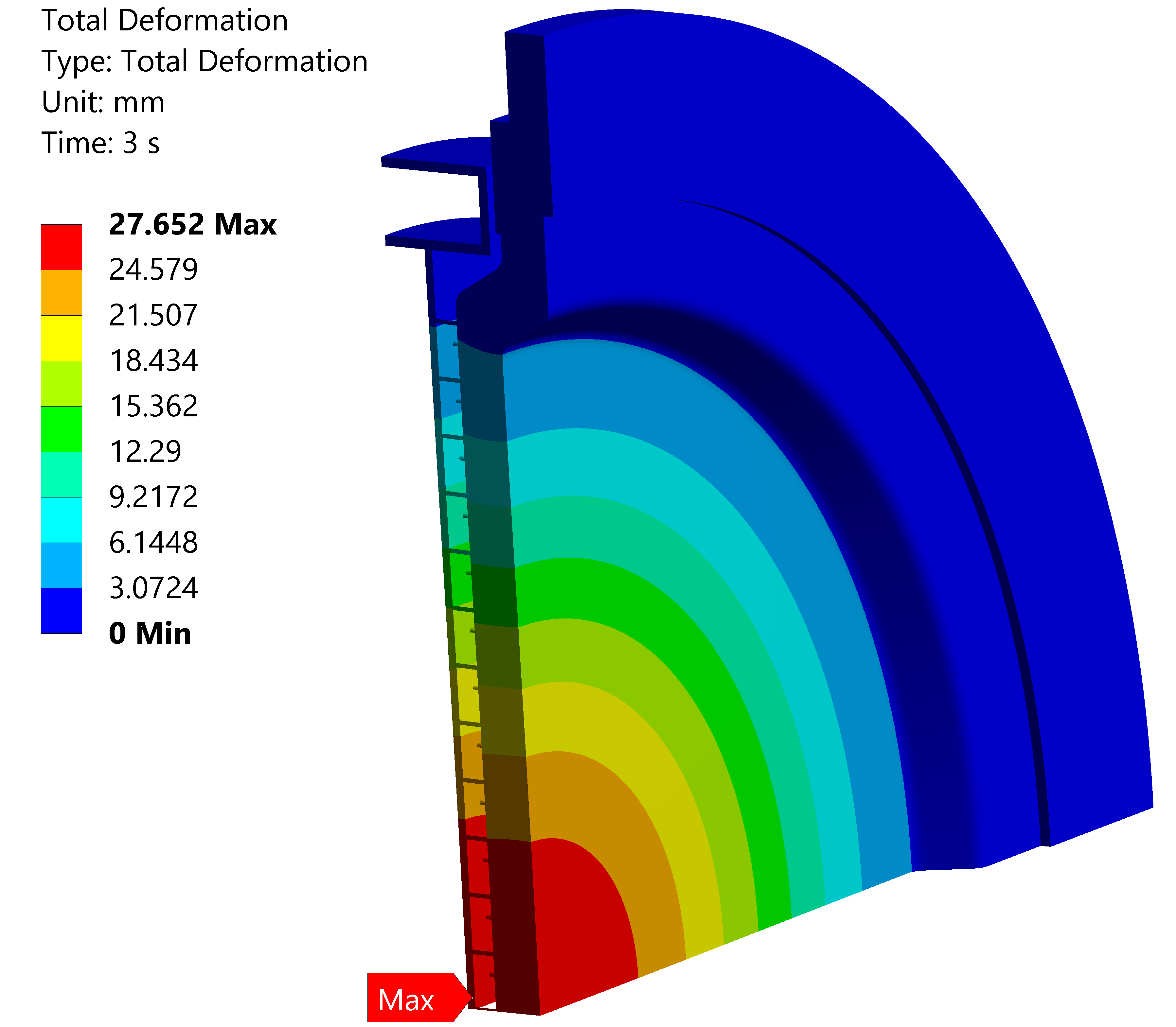}
         \caption{Deformation}
     \label{fig:STAYRESULTS}
     \end{subfigure}
     \begin{subfigure}[b]{0.59\textwidth}
         \centering
         \includegraphics[width=\textwidth]{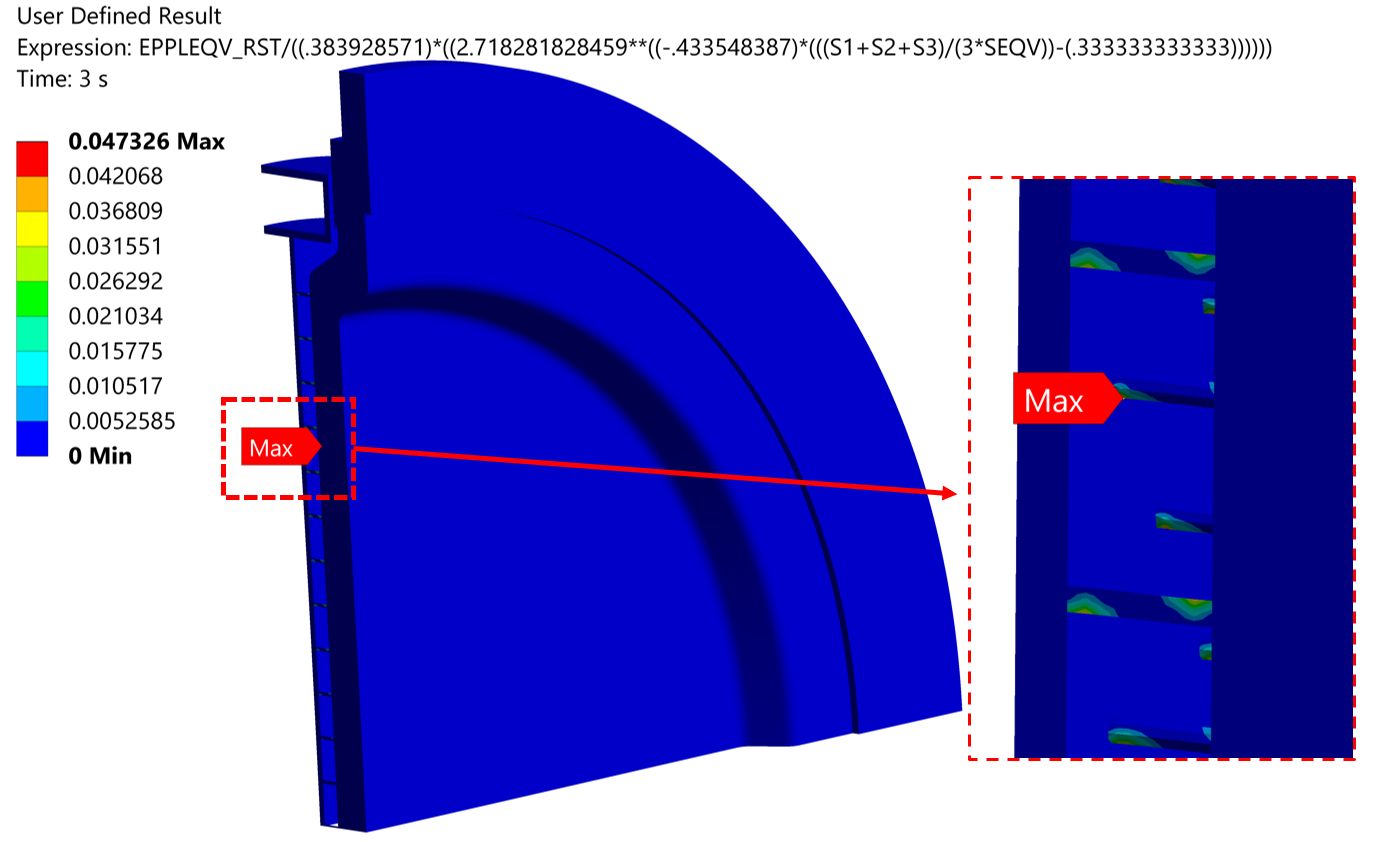}
         \caption{Local failure check}
         \label{fig:STAYRESULTS2}
     \end{subfigure}
        \caption{Simplified stayed head analysis}
        \label{fig:STAYRESULTS3}
\end{figure}

Using the simplified stayed head model, the cover was examined using a 10 bar internal pressure. The results of the local failure and plastic collapse analysis for the analysis model can be seen in Figure~\ref{fig:STAYRESULTS} and Figure~\ref{fig:STAYRESULTS2}.
%
%\begin{figure}[h!]
%\centering
%\includegraphics[width=0.98\textwidth]{Figures/Stayed_Head_Deformation_2.png}
%\caption{The simplified initial stayed head analysis: Deformation}
%\label{fig:STAYRESULTS}
%\end{figure}
%
%\begin{figure}[h!]
%\centering
%\includegraphics[width=0.98\textwidth]{Figures/Stayed_Head_Local_Failure_2_A.png}
%\caption{The simplified initial stayed head analysis: Local failure check}
%\label{fig:STAYRESULTS2}
%\end{figure}
%

The results of this analysis of the stayed head showed that convergence was achieved and the local failure criteria requirements were met meeting the requirements of The Code for this model.

\subsubsection{Pressure vessel/Cryostat shell analysis}

\begin{figure}
     \centering
     \begin{subfigure}[b]{.49\textwidth}
         \centering
         \includegraphics[width=\textwidth]{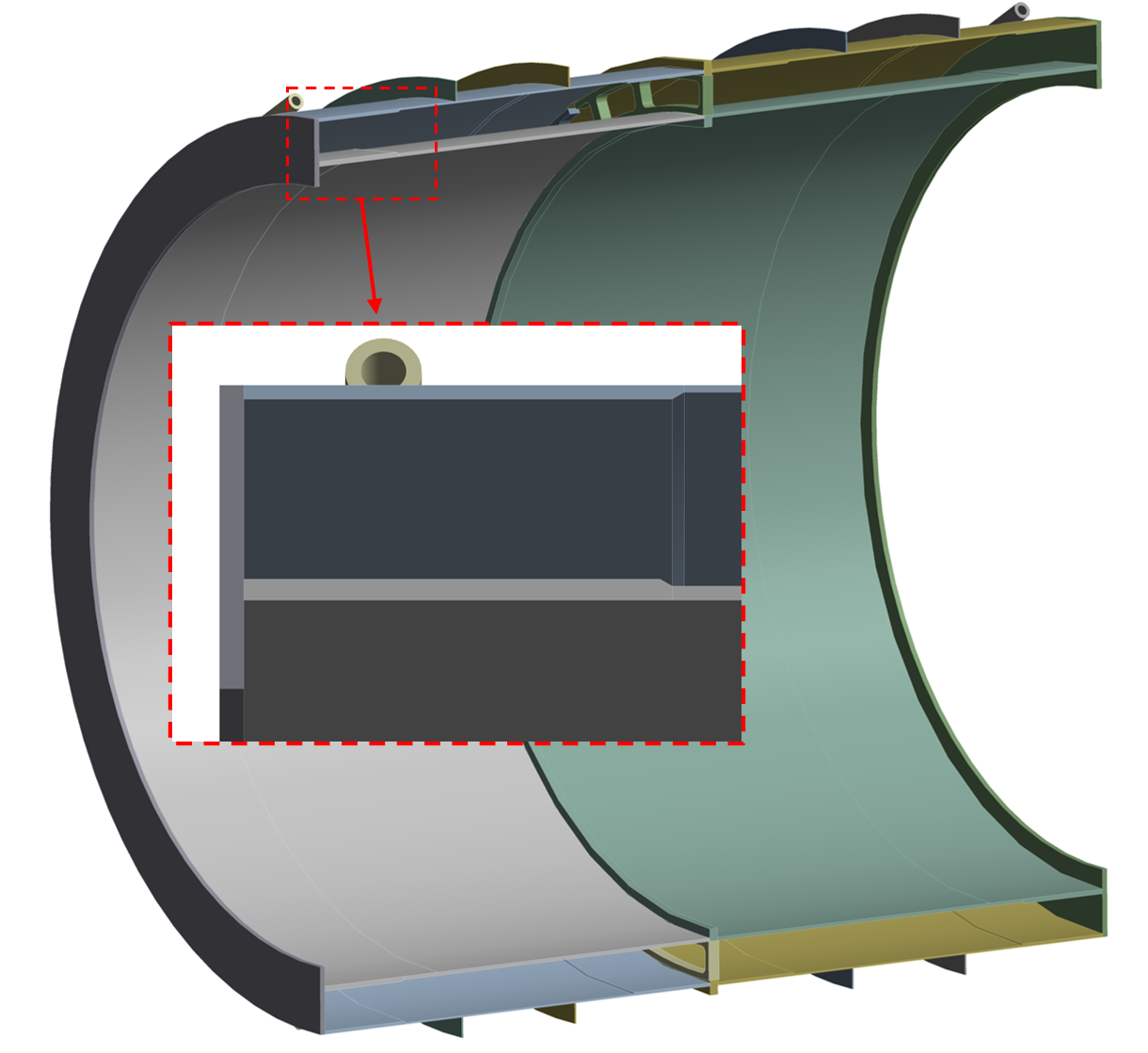}
         \caption{Model used in cryostat shell analysis.}
     \label{fig:CSAM}
     \end{subfigure}
     \begin{subfigure}[b]{.49\textwidth}
         \centering
         \includegraphics[width=\textwidth]{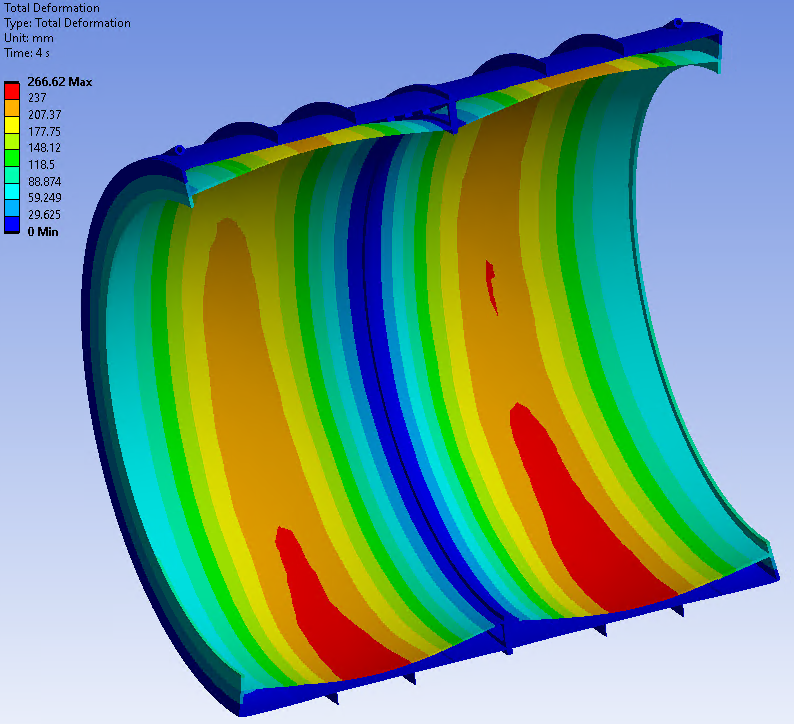}
         \caption{Results of cryostat shell analysis: Plastic collapse requirement: Convergence of analysis model meeting plastic collapse requirements}
         \label{fig:CSAM_Results}
     \end{subfigure}
     \hfill
     \begin{subfigure}[b]{0.42\textwidth}
         \centering
         \includegraphics[width=\textwidth]{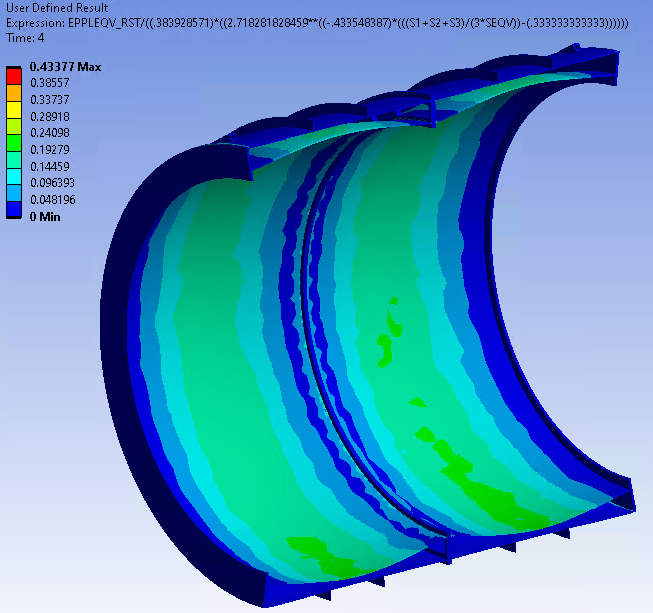}
         \caption{Results of cryostat shell analysis: Local failure requirement.}
         \label{fig:CSAM_Results2}
     \end{subfigure}
     \hfill
     \begin{subfigure}[b]{0.57\textwidth}
         \centering
         \includegraphics[width=\textwidth]{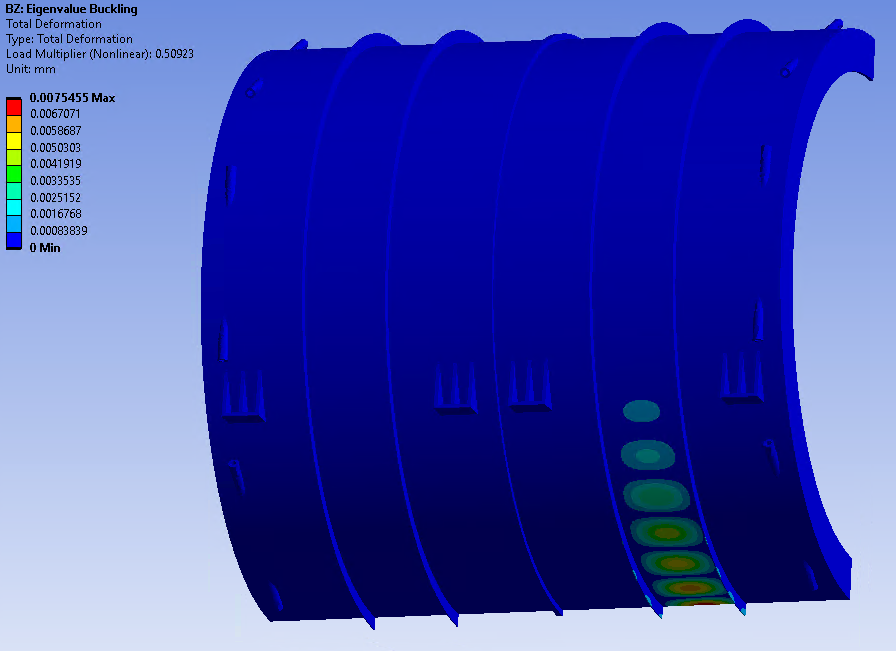}
         \caption{Results of cryostat buckling mode shapes.}
         \label{fig:CSAMBucklingResults}
     \end{subfigure}
        \caption{Cryostat FEA analysis}
        \label{fig:CSAM Results}
\end{figure}

The cryostat shell wall thickness was optimized within the 10 bar pressure constraint. Using elastic-plastic methods as described in The Code, a simplified symmetrical model shown in Figure \ref{fig:CSAM} was evaluated. The model included consideration of the estimated weight of the calorimeter ($\simeq$ 180 t) and TPC ($\simeq$ 15 t) along with the reaction forces from the stayed head assessment and the radial supports. 
%
%\begin{figure}[h!]
%\centering
%\includegraphics[width=0.95\textwidth]{Figures/Analysis_Model_Zoomed_2.png}
%\caption{Model used in cryostat shell analysis.}
%\label{fig:CSAM}
%\end{figure}

The simulation was able to converge on a solution, meeting the protection against plastic collapse requirements. The simulation also shows that the assembly meets the protection against local failure requirement of having a local strain limit, $\epsilon_{limit}$, lower than 1.0. These results can be seen in Figure~\ref{fig:CSAM_Results} and Figure~\ref{fig:CSAM_Results2}. Additional load cases that occur during the manufacturing and transport of the vessel will require a future analysis.
%
%\begin{figure}[h!]
%\centering
%\includegraphics[width=0.98\textwidth]{Figures/Analysis_Model_Collapse.PNG}
%\caption{Results of cryostat shell analysis: Plastic collapse requirement: Convergence of analysis model meeting plastic collapse requirements}
%\label{fig:CSAM_Results}
%\end{figure}
%
%\begin{figure}[h!]
%\centering
%\includegraphics[width=0.98\textwidth]{Figures/Analysis_Model_Local.png}
%\caption{Results of cryostat shell analysis: Local failure requirement.}
%\label{fig:CSAM_Results2}
%\end{figure}
%
The loaded shell from this analysis was then used to determine if the shell would be able to resist buckling. A collapse analysis was performed using imperfections generated through an elastic-plastic buckling review. The buckling mode shapes used to generate a set of imperfections in the model of the Cryostat shell. These imperfections can be seen in Figure~\ref{fig:CSAMBucklingResults}. Using the generated imperfect model, a plastic collapse analysis was performed again to determine if buckling occurred. The model was able to converge on a solution, meeting the protection against buckling requirements.
%
%\begin{figure}[h!]
%\centering
%\includegraphics[width=0.99\textwidth]{Figures/Buckling.png}
%\caption{Results of Cryostat buckling mode shapes.}
%\label{fig:CSAMBucklingResults}
%\end{figure}
%
%\begin{figure}[h!]
%\centering
%\includegraphics[width=0.99\textwidth]{Figures/Buckling2.png}
%\caption{Results of Cryostat Buckling Mode Shapes}
%\label{fig:CSAMBucklingResultsb}
%\end{figure}
%

Before the design of the cryostat can be finished, additional detailing is needed to account for the interaction between the finalized magnet design and detector equipment. Additionally for the final stayed head design aspects such as the attachment methods of the stay bolts, pattern of the connection, and additional failure modes will need to be examined. Despite additional work being needed, the results of the analysis indicate that the design of the cryostat for the superconducting solenoid in SPY is feasible for withstanding operational loading.
\subsection{Pressure vessel head failure mode analyses}
\label{sec:FM}
\begin{figure}[h!]
\centering
\includegraphics[width=0.65\textwidth]{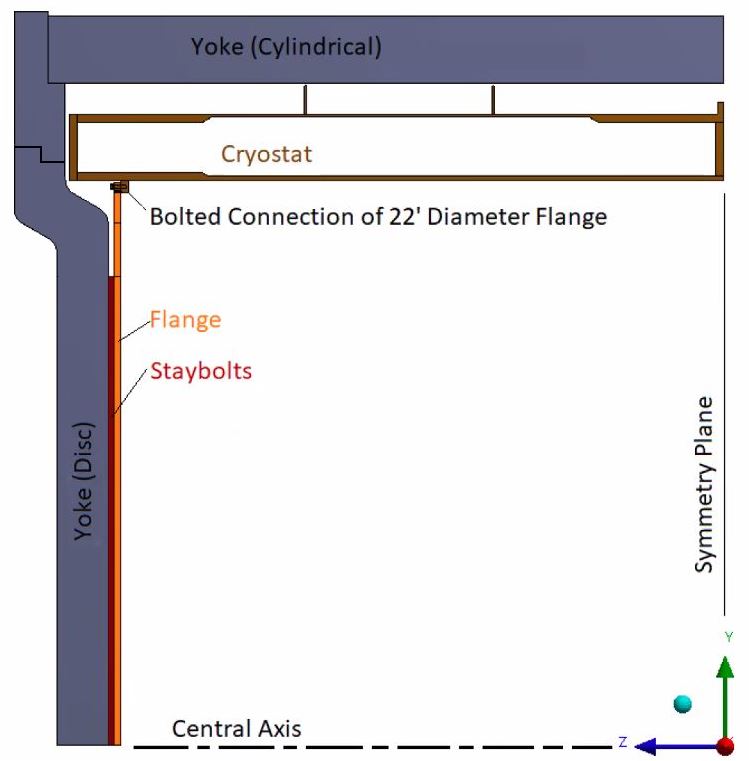}
\caption{Cross section of the cryostat, flange, and yoke, where the central axis of the cylinder is the z-axis.}
\label{fig:FlangeCrossSection}
\end{figure}

The SPY pressure containment system utilizes a very large flat circular flange (6.6m in diameter) which is designed to hold 10 bar of gaseous argon.  The flange and flange bolts themselves are not strong enough to support the pressure (4kt of force) over such a large area.  This force will be contained by the yoke as described in Section~\ref{sec:PVHA}.
Figure~\ref{fig:FlangeCrossSection} shows a cross section of the cryostat, flange, and yoke, where the central axis of the cylinder is the z-axis, and the bolted flange connection of interest is called out.
%

%\begin{figure}[h!]
%\centering
%\includegraphics[width=0.95\textwidth]{Figures/Radial Pressure.png}
%\caption{Radial pressure on the flange from the stay bolts vs. radius when a traditional flat face bolting method is used.}
%\label{fig:radialPressure}
%\end{figure}

\begin{figure}[h!]
\centering
\includegraphics[width=0.50\textwidth]{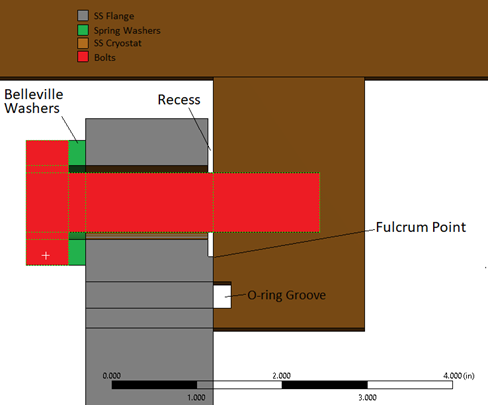}
\caption{Flexible bolted flange connection allows rotation of the flange end, eliminates the prying force and moment on the bolts, and keeps the vessel sealed.}
\label{fig:FlexBoltedFlange}
\end{figure}

We have developed a bolting method for the flange which is a simply-supported type connection.  This flexible bolted connection eliminates the bending moment on the bolts, as it allows rotation about the new fulcrum point, which is placed near the O-ring groove instead of at the outer edge of the flange.  Since it is more flexible, it also results in more of the total force being transferred to the yoke, as opposed to the flange itself.  This connection method is achieved by using a machined recess in the flange as well as spring washers of an appropriate spring constant with and large deformation range. Figure~\ref{fig:FlexBoltedFlange} shows the flange connection.  Bolt pre-tension must be high enough to keep the fulcrum point on the flange in contact with the vessel's mating flange.  The analysis shows this flexible flange connection is able to keep the flange sealed, and has adequate bolt strength.  Choosing the correct spring washers with additional working deformation, along with the appropriate pre-tension, will be of vital importance in making this connection work as intended.

\subsection{Yoke}
\label{subsec:Yoke}
\begin{figure}[h!]
\centering
\includegraphics[width=0.65\textwidth]{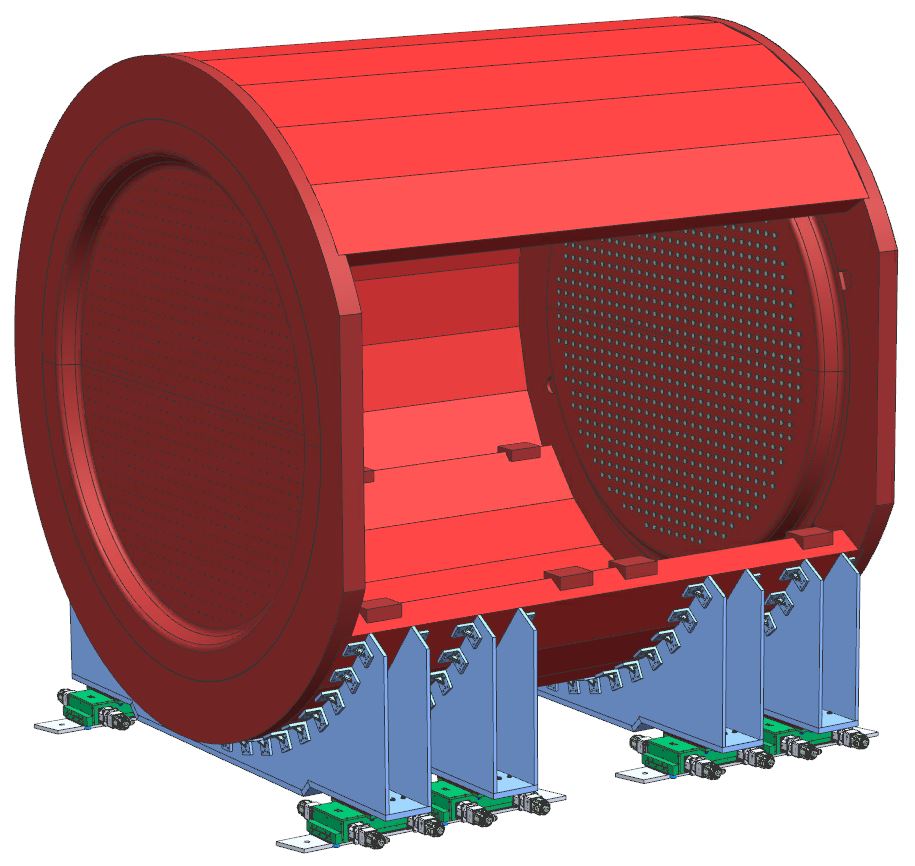}
\caption{Full Yoke w/o solenoid.}
\label{fig:Full_Yoke}
\end{figure}

The steel yoke and cradles of SPY (see Figure~\ref{fig:Full_Yoke}) weigh 881 tons. The yoke must be fabricated from low carbon steel in order to contain the magnetic flux. The four cradles do not contribute to the development of the magnetic field and will be either 18-8 or 304 stainless steel. Each component of the yoke system will be under the crane limit of 60 tons and will fit within the constraints of the access shaft. The four end plug sections are the heaviest pieces of the yoke assembly, each weighing 54.5 tons. There are two end rings and each weighs 46.3 tons. The long, axial steel plates that make up the barrel of the yoke come in two thicknesses, 150 mm and 400 mm. The 400 mm thick plates each weigh 26 tons. The cradles each weigh 10.7 tons.

The steel axial plates will be bolted to the four cradles and after the solenoid is installed, the remaining yoke components will be assembled via a bolted construction. Finite element analysis results show that the steel under gravity loading has minimal deflection but under magnetic loading and creep conditions, begins to deform. The yoke will need a constraint system of either bolts, welds, or straps to preserve the alignment of the separate components.  Once the end plugs are installed, the stay bolts must be tightened to provide contact to the pressure containment end flanges. The infrastructure needs of the magnet system will impact the yoke design.  These include the cables and piping that will need to pass through the yoke and instrumentation inside the yoke for monitoring experimental parameters and the yoke itself.  Further refinement and optimization of the yoke assembly will need to be performed as the cryostat design matures.
\subsubsection{Movement system}
\label{sec:MS}
\begin{figure}[h!]
\centering
\includegraphics[width=0.65\textwidth]{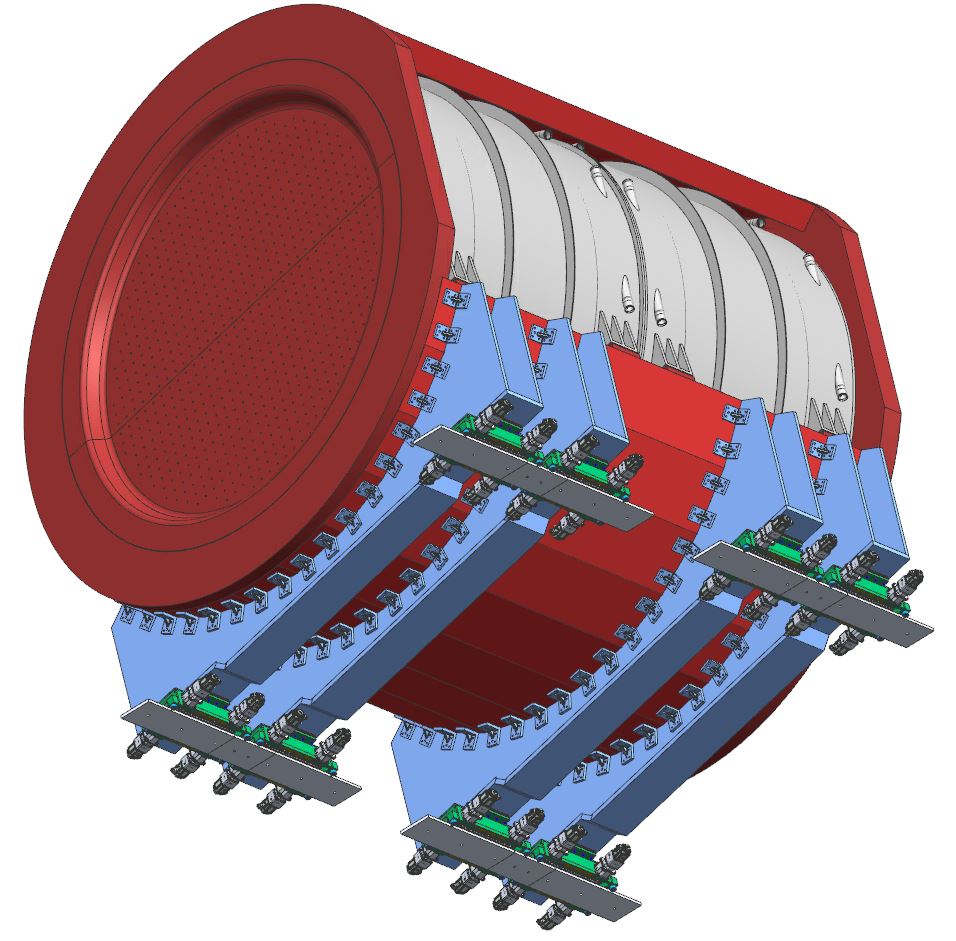}
\caption{Motorized industrial rollers and support cradles.}
\label{fig:Stayed-head}
\end{figure}

The motorized Hilman roller system~\cite{site:Hilman} that will transport ND-GAr is shown in Figure~\ref{fig:Stayed-head}. Eight, 200 ton capacity Hilman rollers have been specified for this application. Each roller is motorized and will be operated in series to transport the magnet system along steel flat tracks to the required hall positions.

\newpage
\subsubsection{Yoke FEA analysis}

\begin{figure}
     \centering
     \begin{subfigure}[b]{.47\textwidth}
         \centering
         \includegraphics[width=\textwidth]{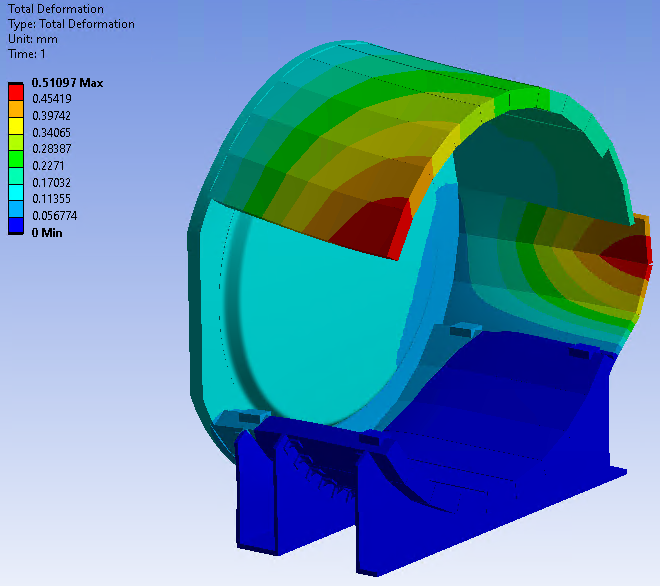}
         \caption{Deformation of the Yoke.  The red zones indicate a deformation of slightly over one-half of a mm.}
     \label{fig:Yoke Deform}
     \end{subfigure}
     \begin{subfigure}[b]{.50\textwidth}
         \centering
         \includegraphics[width=\textwidth]{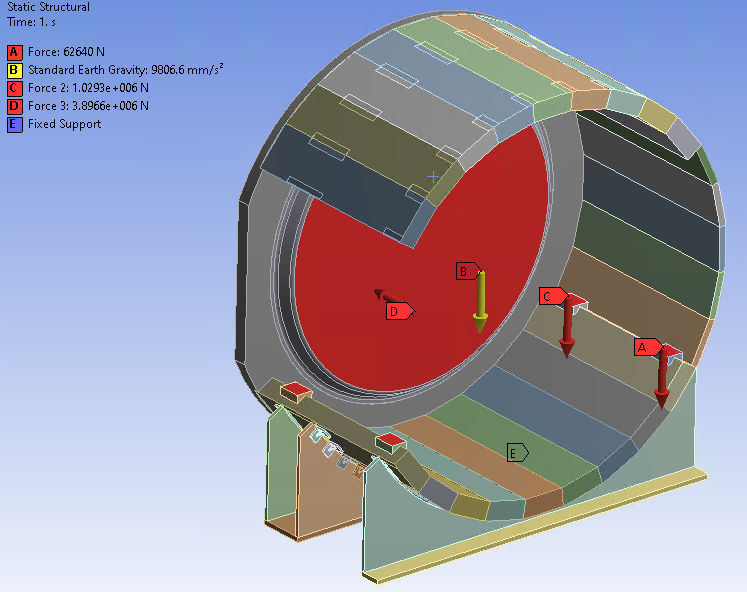}
         \caption{Boundary conditions of yoke structural simulations.}
         \label{fig:Yoke_BC}
     \end{subfigure}
     \hfill
     \begin{subfigure}[b]{0.49\textwidth}
         \centering
         \includegraphics[width=\textwidth]{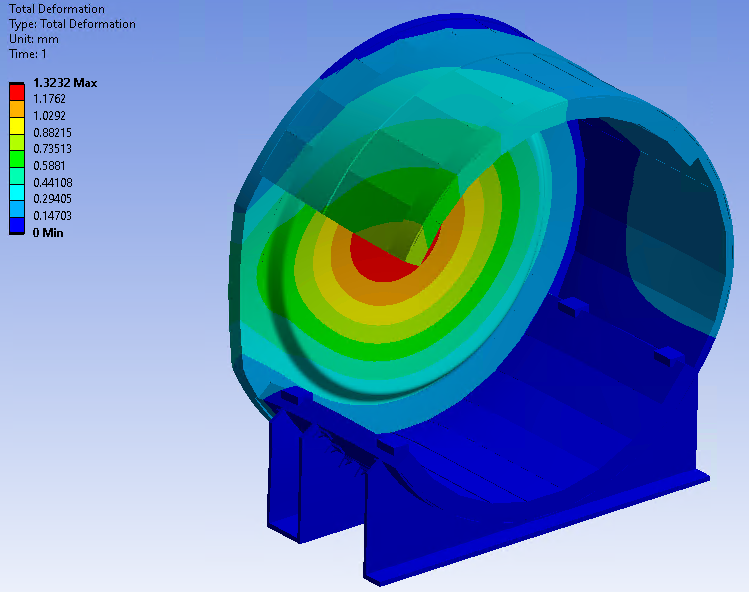}
         \caption{Results of hoop yoke analysis: Total deformation.}
         \label{fig:Hoop Yoke Deform}
     \end{subfigure}
     \hfill
     \begin{subfigure}[b]{0.48\textwidth}
         \centering
         \includegraphics[width=\textwidth]{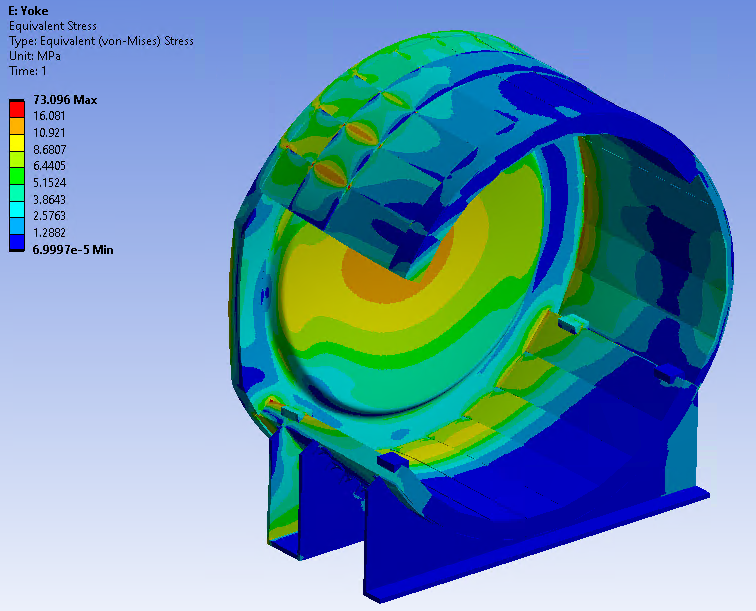}
         \caption{Results of hoop yoke analysis: Equivalent stress.}
         \label{fig:Hoop Yoke Stress}
     \end{subfigure}
        \caption{Yoke FEA Analysis}
        \label{fig:Yoke FEA Analysis Results}
\end{figure}

To determine if the yoke will be able to support and absorb the loads of the assembly, a simple FEA analysis using reaction forces from the FEA analyses of the pressure-vessel head and the cryostat/pressure-vessel shell was performed. Using only the gravitational loads of the yoke and solenoid, the initial results shown in Figure~\ref{fig:Yoke Deform} indicate that unrestrained  sections of the yoke will begin slipping. To ensure stability of the assembly, the individual yoke sections will need to be bound together with hoop sections.
%
%\begin{figure}[b]
%\centering
%\includegraphics[width=0.98\textwidth]{Figures/Yoke_Deformation.PNG}
%\caption{Deformation of the Yoke.  The red zones indicate a deformation of slightly over one-half of a mm.}
%\label{fig:Yoke Deform}
%\end{figure}
%
%
The hoops help to evenly distribute the gravitational loads and make the yoke assembly sufficiently strong. Additionally, the results indicate that the assembly is sufficiently strong. The boundary conditions of the simulations can be seen in Figure~\ref{fig:Yoke_BC}. The stress and deformation results of the simulation can be seen in  Figure~\ref{fig:Hoop Yoke Deform}, and Figure~\ref{fig:Hoop Yoke Stress}. The results show that the deformation is negligible and the stresses in the model are manageable. Due to simplifications of the assembly FEA model, non-realistic stresses develop at localized regions. Further refinement of the design and FEA model is required to finalize the yoke design.

%\begin{figure}[h]
%\centering
%\includegraphics[width=0.98\textwidth]{Figures/Yoke_BC.PNG}
%\caption{Boundary conditions of yoke structural simulations.}
%\label{fig:Yoke BC}
%\end{figure}

%\begin{figure}[h]
%\centering
%\includegraphics[width=0.98\textwidth]{Figures/Hoop_Yoke_Deformation.PNG}
%\caption{Results of hoop yoke analysis: Total deformation.}
%\label{fig:Hoop Yoke Deform}
%\end{figure}
%
%\begin{figure}[h]
%\centering
%\includegraphics[width=0.98\textwidth]{Figures/Hoop_Yoke_Stress.PNG}
%\caption{Results of hoop yoke analysis: Equivalent stress.}
%\label{fig:Hoop Yoke Stress}
%\end{figure}
%

\clearpage
\section{Preliminary parameter set}
In Table~\ref{tab:M_req} we list the preliminary parameter set for the SPY magnet system.
\begin{table}[h]
\small
\begin{center}
\caption{Preliminary parameter set for SPY magnet system.}
\label{tab:M_req}
\begin{tabular}{|l|c|c|}
\hline
\textbf{Parameter} &  \textbf{Value} & \textbf{Notes }\\ \hline
\hline
Central field & 0.5T &\\ \hline
Field uniformity & $\pm$ 10\% & Current design achieves $\pm$ 2\%\\ \hline
Ramp time to full field & 30 min & \\ \hline
Stray field & $\leq$ 0.01 T & Stray field in SAND negligible, in LAr FV $\simeq$ 10G \\ \hline
Bore & 6.725m & Reduction possible with HPgTPC and ECAL optimization\\ \hline
OD & 7.85m & Cryostat diameter at stiffening rings\\ \hline
Length & $\simeq$ 7.8m & Cryostat length\\ \hline
Solenoid weight & $\simeq$ 150t & \\ \hline
Yoke total weight & $\simeq$ 757t & \\ \hline
\end{tabular}
\end{center}
\end{table}

\section{Conclusion}

A magnetic and mechanical conceptual design for the magnet system for a high pressure, gaseous argon neutrino detector has been developed. This design relies on the experience on numerous magnets built over the past decades, but features several unique characteristics. Its bore would make it the largest superconducting magnet ever used in particle physics and the requirement for a low material budget for the solenoid is reflected in a thin solenoid design. The iron yoke is asymmetric to allow for particles that enter from an upstream detector to be tracked in the HPgTPC in the magnet's bore. Finally, the integration of the detector and the solenoid cryostat is complete, using the inner shell of the magnet vacuum chamber as the outer shell for high pressure containment for the HPgTPC. To limit the overall length of the assembly, the pressure is also transferred to the iron yoke end caps with a dedicated system of stay bolts, thus accommodating the use of thin, flat end flanges for gas containment.

%
%
%\section{Instrumentation}
%%%%    \item Voltage taps
%    \item Strain gauges
%    \item Pressure and vacuum gauges
%\end{enumerate}
%
%Since the vacuum vessel will also be used to contain the 10 bar pressure volume containming the HPgTPC and ECAL, additional strain gauges on the inner and outer shells are envisioned.
%
%\subsection{In yoke}
%Instrumentation from within the detector will pass through electrical feed-through plates mounted along the fixed flange of the pressure vessel so that the large end flange can be removed for repairs or maintenance without the need to disconnect the DAQ cables. The cables will then be routed around or through the magnet yoke by means of penetrations in the steel or by routing to the large yoke opening between ND-GAr and ND-LAr and then on to cable trays mounted alongside of the flexible cryogenic lines.
%
%\section{Detector interfaces}
%
%The solenoid cryostat will need to carry the load of the HPgTPC and the ECAL.  See Figure ~\ref{fig:ECAL} where the ECAL is shown in blue and the HPgTPC in yellow.  The total mass the two detectors will be less than 150t with the majority of the mass being the ECAL.  Optimization of studies of the ECAL will likely yield a much lighter detector, but final conclusions have not yet been reached.
%
%\begin{figure}[t]
%\centering
%\includegraphics[width=0.90\textwidth]{Figures/ECAL+HPgTPC.png}
%\caption{Schematic of ECAL and HPgTPC mounting to solenoid}
%\label{fig:ECAL}
%\end{figure}
%
%\newpage
%
\clearpage
\bibliographystyle{JHEP}
\bibliography{references}

\end{document}